\documentclass[useAMS,usenatbib]{mn2e}
\usepackage{amsmath}
\usepackage{amssymb}
\usepackage{graphicx}
\usepackage{txfonts}
\usepackage{journals}
\usepackage{wasysym}
\usepackage{xcolor}
\usepackage{todonotes}
\usepackage{verbatim}
\title[Two Modes of Partially Screened Gap]{Two Modes of Partially Screened Gap }

\author[Szary et al.]{
  A. Szary$^{1}$\thanks{E-mail: aszary@astro.ia.uz.zgora.pl}, G. Melikidze$^{1,2}$ and J. Gil$^{1}$ \\
  $^1$ Kepler Institute of Astronomy, University of Zielona G\'ora , Lubuska 2, 65-265 Zielona G\'ora, Poland\\
  $^2$ Ilia State University, E. Kharadze Abastumani Astrophysical Observatory, Tbilisi, Georgia}
  
\begin{document}

%\date{Accepted 1988 December 15. Received 1988 December 14; in original form 1988 October 11}
\date{Accepted . Received ; in original form }

\pagerange{\pageref{firstpage}--\pageref{lastpage}} \pubyear{2013}
\maketitle
\label{firstpage}

\begin{abstract}
	The analysis of X-ray observations suggest an ultrastrong ($B\gtrsim 10^{14} \,{\rm G}$) surface magnetic field at the polar cap of pulsars \citep{2013_Szary}.
	On the other hand, the temperature of the polar caps is about a few millions Kelvin.
	Based on these two facts we use the Partially Screened Gap (PSG) model to describe the Inner Acceleration Region (IAR).
	The PSG model assumes that the temperature of the actual polar cap is equal to the so-called critical value, i.e. the temperature at which the outflow of thermal ions from the surface screens the gap completely.
	We have found that, depending on the conditions above the polar cap, the generation of high energetic photons in IAR can be caused either by Curvature Radiation (CR) or by Inverse Compton Scattering (ICS).
	Completely different properties of both processes result in two different scenarios of breaking the acceleration gap: the so-called PSG-off mode for the gap dominated by CR and the PSG-on mode for the gap dominated by ICS.
	The existence of two different mechanisms of gap breakdown naturally explains the mode-changing phenomenon.
	Different characteristics of plasma generated in the acceleration region for both modes also explain the pulse nulling phenomenon.
\end{abstract}

\begin{keywords}
    {(stars:) pulsars: general -- radiation mechanism: non-thermal -- etc.}
\end{keywords}

\section{Introduction}
	
% start from here

The classical vacuum gap model of inner acceleration region was proposed by \cite{1975_Ruderman} assuming that the conditions just above the polar cap are near to a pure vacuum. Assuming the dipolar structure of the magnetic field they estimated the polar cap area as \mbox{$A_{\rm dp}\approx 6.2 \times 10^4 P^{-1} \, {\rm m^2}$} and the magnetic field strength at the stellar surface as \mbox{$B_{\rm d}=2.02 \times 10^{12} (P \dot{P}_{-15} )^{0.5}$}, where $P$ is the rotation period, $\dot{P}_{-15}=10^{15}\dot{P}$ and  $\dot{P}$ is the period time derivative. On the other hand \cite{1975_Ruderman} applied a highly non-dipolar radius of curvature of field lines ($\Re \approx 10^6 \,{\rm cm}$) in order to enable an effective electron-positron cascade. It is widely accepted that pulsar activity is conditioned by the efficiency of pair creation and it defines the theoretical death line $\Phi_{\rm death} \approx 10^{13} (\dot{P} / 10^{-15})^{0.5} P^{-1.5} \,{\rm V}\lesssim 10^{12} \, {\rm V}$  (assuming the magnetic field $B\approx B_{\rm d}$) \citep{1975_Ruderman, 1992_Bhattacharya, 1993_Chen}. As it was shown by \cite{2009_Arons}, the theoretical death line reasonably well reproduces the observed sample (see their Figure 15.1).  Recently, however, it was shown by \cite{2014_Szary} that the observed sample  can be reproduced much better by imposing the limit on radio efficiency $\xi=L/\dot{E}<0.01$, where $L$ is radio luminosity, and $\dot{E}$ is called spin-down luminosity, the rate of loss of rotational kinetic energy.  In this approach, a few pulsars can still be detectable in the graveyard region as seen in real observations.  
Furthermore, it was shown that the random luminosity model, which was discarded in previous studies \citep[see, e.g.][]{2006_Faucher}, is both justified by the weak observational correlations and results in much better fit to the observed sample.
The fact that both the location of death line and radio luminosity are not a function of either $P$ or $\dot{P}$ suggest that the plasma responsible for radio emission is generated in a region where magnetic field is no longer a simple function of $P$ and $\dot{P}$, and thus $B \not \approx B_{\rm d}$.
    
    Many authors studied the pair creation initiated by energetic particles moving in the magnetosphere above the polar cap within the framework of the vacuum gap \citep[e.g.,][]{1975_Ruderman, 1980_Cheng, 1997_Zhang_b, 2002_Gil_b, 2006_Timokhin, 2010_Medin}.
    Influence of the non-dipolar configuration of the surface magnetic field on the plasma production was postulated from the very beginning of the pulsar theory \citep[see, e.g.,][]{1971_Sturrock, 1975_Ruderman, 1980_Jones}, however, it  was barely explored.
    At the same time, there is increasing evidence (both observational and theoretical) that the surface magnetic field is non-dipolar. For instance, the X-ray observations of old pulsars show that an actual polar cap size, $\mathbf{A_{\rm bb}}$, is much smaller than the one that follows from a purely dipole field structure, $A_{\rm bb} \ll A_{\rm dp}$ \citep[see][and references therein]{2009_Becker, 2013_Szary}. Furthermore, \cite{2013_Geppert} found that the crustal Hall drift can produce very strong and small-scale surface magnetic field structures by means of non-linear interaction between poloidal and toroidal components of the subsurface magnetic field.  
In order to  overcome weaknesses of the vacuum gap model, \cite{2003_Gil} incorporate the non-dipolar geometry of the crustal magnetic field and obtain naturally a proper rate of the subpulse drift. 
The Partially Screened Gap (PSG) model was introduced, which allows to study the most intriguing region around a pulsar, namely the inner acceleration region.
    
In this paper, we study the possible scenarios of PSG breakdown. In Section \ref{sec:psg} we present the details of the model. A method to define main parameters of PSG and possible scenarios of gap breakdown are presented in Section \ref{sec:results}. The results are summarised and discussed in Section \ref{sec:discussion}.  

\section{Partially Screened Gap}\label{sec:psg}

The charge-depleted inner acceleration region above the polar cap can be formed if a local charge density differs from the co-rotational charge density $\rho_{{\rm GJ}}={\bf \Omega}\cdot{\bf B}/\left(2\pi c\right)$ \citep{1969_Goldreich}. The sign of the charge depends on the mutual orientation of the neutron's star rotation axis ${\bf \Omega}$ and magnetic moment $\bf B$: the sign is positive  if ${\bf \Omega}\cdot{\bf B}<0$ and negative if ${\bf \Omega}\cdot\ {\bf B}>0$. Therefore, the charge depletion above the polar cap depends on the binding energy of either the positive ${\rm _{26}^{56}Fe}$ ions (${\bf \Omega}\cdot{\bf B}<0$) or electrons (${\bf \Omega}\cdot{\bf B}>0$).
\cite{2007_Medin} showed that electrons can easily escape from the condensed surface, preventing the formation of an acceleration region just above the polar cap (both the vacuum gap and PSG).
In such a scenario the density of these electrons just above the stellar surface is high enough to completely screen the acceleration potential.
In the case of negatively charged polar caps other acceleration mechanisms can work, for instance the space-charged-limited flow \citep{1979_Arons}, or the slot gap \cite{1983_Arons}.
However, these mechanisms are based on the fact that charge depleted regions are formed at heights much above the stellar surface, where the formation of PSG is not possible due to lack of source of ions which could screen the acceleration potential.
Therefore, in this paper we consider only the case of positively charged polar caps (${\bf \Omega}\cdot {\bf B}<0$). We assume that the crust of neutron stars mainly consists of iron $\left({\rm _{26}^{56}Fe}\right)$ formed at the birth of a star \cite[e.g.,][]{2001_Lai}.	Emission of charged particles (iron ions) from the condensed stellar surface depends on its temperature. When a cohesive energy of iron ions is high enough, the positive charges cannot be supplied at a rate that would compensate their inertial outflow through the light cylinder \citep[see][]{2006_Medin_a, 2006_Medin_b, 2007_Medin,2007_Gil_b}. This is actually possible if the surface temperature $T_{{\rm s}}$ is below the so-called critical value $T_{{\rm crit}}$.
The critical temperature is the temperature at which the charge density of ions emitted from the surface is equal to the co-rotational charge density $\rho_{\rm i} =  \rho_{\rm GJ}$. Since the density of the iron ions in the neutron star crust is many orders of magnitude larger than the co-rotational charge density, a thermionic emission from the polar cap surface is not simply described by the usual condition $\epsilon_{{\rm i}}\approx kT_{{\rm s}}$, where $\epsilon_{i}$ is the cohesive energy and/or work function, $T_{{\rm s}}$ is the actual surface temperature, and $k$ is the Boltzman constant.	The outflow of iron ions can be described in the form \citep[see][and references therein]{2003_Gil}
		\begin{equation}
    		\frac{\rho_{{\rm i}}}{\rho_{{\rm GJ}}}\approx \exp \left(C_{{\rm i}}-\frac{\epsilon_{i}}{kT_{{\rm s}}}\right),
		\end{equation}
where $\rho_{{\rm i}}\leq\rho_{{\rm GJ}}$ is the charge density of the outflowing ions.

As soon as the surface temperature $T_{{\rm s}}$ reaches the critical value
		\begin{equation}
    		T_{{\rm crit}}=\frac{\epsilon_{i}}{C_{{\rm i}}k},
		\end{equation}
the ion outflow reaches the maximum value ($\rho_{{\rm i}}=\rho_{{\rm GJ}}$).	The numerical coefficient $C_{{\rm i}}=30\pm 3$ is determined from the tail of the exponential function with an accuracy of about 10\%. The cohesive energy of condensed matter is mainly defined by the strength of the magnetic field and was calculated by \cite{2006_Medin_a, 2006_Medin_b, 2007_Medin}. Their calculations indicate that the cohesive energy increases with magnetic field strength and becomes significant at $B_{\rm s} \gtrsim 10^{14}\,{\rm G}$. Uncertainties in calculation of the cohesive energy make it known to within a factor of two \citep{2006_Medin_a, 2006_Medin_b}. Thus, for a given value of the cohesive energy, the critical temperature $T_{{\rm crit}}$ is also estimated within a factor of two accuracy \citep{2007_Medin}. Furthermore, calculations for $B_{\rm s}<10^{14}\,{\rm G}$ should be treated with caution, because in this regime the bound iron molecules are not decidedly more favourable than isolated atoms.
    
    \subsection{The model}
    
        \begin{figure}
        \begin{center}
                \includegraphics[width=7.5cm]{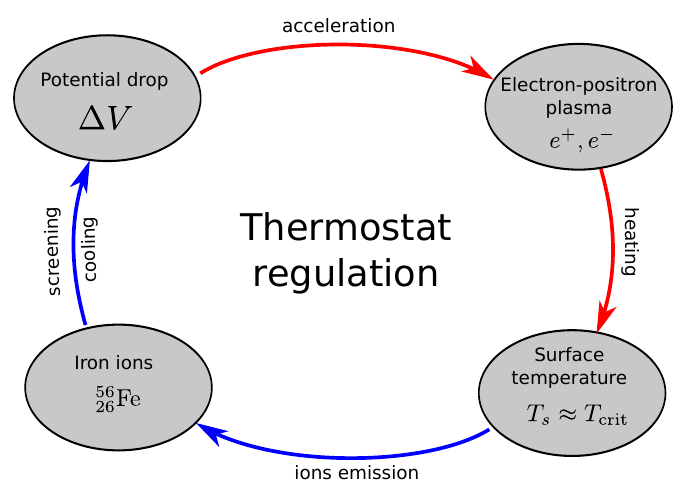}
        \end{center}
        \caption{Thermostatic regulation of the polar cap temperature in the PSG model. The surface temperature at the polar cap is thermostatically self-regulated within a narrow range around the critical temperature $T_{\rm crit}$.}
                \label{fig:thermostat}
        \end{figure}

As it follows from the X-ray observations \citep[see, e.g.,][and references therein]{2009_Becker, 2013_Szary}, the temperature of the hot spot (which is associated with the actual polar cap) is about a few times $10^{6}\,{\rm K}$. In order to sustain such a high temperature the bombardment by the backstreaming particles is required.	However, particle acceleration (and therefore the surface heating) is possible only if $T_{{\rm s}}<T_{{\rm crit}}$. \cite{2003_Gil} introduced the Partially Screened Gap model to describe the polar gap sparking discharge specifically under such circumstances.

The PSG model assumes the presence of the iron ions (${\rm _{26}^{56}Fe}$) just above the polar cap with a charge density near, but still below the co-rotational charge density ($\rho_{{\rm GJ}}$). The thermal ejection of ions from the surface causes partial screening of the acceleration potential drop. The degree of screening can be described by the screening factor
		\begin{equation}
    		\eta=1-\rho_{{\rm i}}/\rho_{{\rm GJ}}.
		\end{equation}
Thus, the potential drop can be written as follows:
		\begin{equation}
    		\Delta V=\eta\Delta V_{{\rm max}},\label{eq:deltaV}
		\end{equation}
where $\Delta V_{{\rm max}}$ is the potential drop in a vacuum gap.	We can express the dependence of the critical temperature on the pulsar parameters by fitting to the numerical calculations of \citet{2007_Medin}		\begin{equation}
    		T_{{\rm crit}}\approx 2 \times10^{6} \, B_{14}^{0.75} \, {\rm K}, \label{eq:t_crit}
		\end{equation}where $B_{14}=B_{{\rm s}}/\left(10^{14}\,{\rm G}\right)$, $B_{{\rm s}}=bB_{{\rm d}}$ is a surface magnetic field, and $b=A_{\rm dp}/A_{\rm bb}$ (applicable only if the black-body component is observed, i.e. $A_{\rm bb} < A_{\rm dp}$).	From one hand, the potential drop leads to a polar cap heating due to acceleration of electrons towards the stellar surface.  On the other hand when heated surface reaches temperature close to (or higher than) the critical temperature then emitted ions decrease (or completely eliminates) particle acceleration, and thereby polar cap heating. 
    In a sense PSG works as a typical thermostat in which the same mechanism (acceleration of charge particles in the gap) is responsible for both heating and cooling of the polar cap (see Figure \ref{fig:thermostat}).

	\subsection{Acceleration potential drop}

	As the actual polar cap is much smaller than the conventional (dipolar) polar cap \citep[see][]{2013_Szary}, we cannot use the approximation proposed by \citet{1975_Ruderman} that the gap height is of the same order as the gap width ($h\approx h_{\perp}$).
	On the contrary, the small polar cap size and subpulse phenomenon suggest that the spark half-width is smaller than the gap height ($h_{\perp} < h$).
	For such a regime we have recalculated a formula for the acceleration potential drop $\Delta V$.
	We can estimate the potential drop in a spark region
		\begin{equation}
    		\frac{\Delta V}{h^{2}}+\frac{\Delta V}{h_{\perp}^{2}}=\frac{2\eta B_{\rm s}\Omega\cos\left(\alpha+\vartheta\right)}{c},\label{eq:deltaV_hperp}
		\end{equation}
    where $\alpha$ is the inclination angle between the rotation and the magnetic axis, $\vartheta$ is an angle describing position at the polar cap.
	If we use the same assumptions as \cite{1975_Ruderman}, i.e.: (1) the spark half-width is of the same order as the gap height $h_{\perp}=h$, (2) there is no ion extraction from the stellar surface ($\eta=1$), and (3) the pulsar magnetic and rotation axes are aligned ($\alpha=0^{\circ}$) we get: 
		\begin{equation}
    		\Delta V_{{\rm RS}}=\frac{B_{s}\Omega}{c}h^{2}.
		\end{equation}
	Note that the potential drop defined by Equation \ref{eq:deltaV_hperp} differs from that used by \cite{1975_Ruderman} by the screening factor (as the presence of ions screens the gap), and by the factor of $\cos\left(\alpha+\vartheta\right)$ which also takes into account non-aligned pulsars.
	In our case the polar cap size is much smaller than the conventional polar cap size.
	It seems reasonable to consider sparks with widths much smaller than the gap height ($h_{\perp}\ll h$), furthermore, even for a relatively small inclination angle between the rotation and magnetic axes, we can still write $\vartheta\ll\alpha$.
	In this regime the potential drop can be calculated as
		\begin{equation}
    		\Delta V=\frac{4\pi\eta B_{s}\cos\alpha}{cP}h_{\perp}^{2}.\label{eq:potential_drop}
		\end{equation}

	Since the exact dependence of the electric field on distance above the stellar surface, $z$, is unknown we use the same linear approximation as \cite{1975_Ruderman}.
	In the frame of the PSG model as $h_{\perp}<h$ or even $h_{\perp}\ll h$, we can use a zeroth approximation $\left\langle h_{\perp}E_{\theta}\right\rangle =\left\langle hE_{r}\right\rangle =\Delta V$ and Equation \ref{eq:potential_drop} to describe the component of the electric field along the magnetic field line as follows:
		\begin{equation}
    		E\approx\frac{8\pi\eta B_{{\rm s}}\cos\alpha}{cP}\frac{h_{\perp}^{2}}{h^{2}}\left(h-z\right), \label{eq:acceleration_field}
		\end{equation}
	which vanishes at the top $z=h$.
	The Lorentz factor of a particle after passing distance $l_{{\rm acc}}$ can be calculated as:
		\begin{equation}
    		\gamma=\frac{e}{mc^{2}}\int_{z_{1}}^{z_{2}} Edz\approx\frac{8\pi\eta B_{{\rm s}}e\cos\alpha}{mc^{3}P} \frac{h_{\perp}^{2}}{h^{2}} \left(z_{2}-z_{1} \right) \left(h-\frac{z_{1}+z_{2}}{2}\right),
    		\label{eq:gamma_z}
		\end{equation}
	here $e$ - the electron charge, and $z_{2}-z_{1}=l_{{\rm acc}}$.
	Assuming that a non-relativistic particle is accelerated from the stellar surface ($z_{1}=0$, $\gamma_{0}=1$) we can calculate the distance $l_{{\rm acc}}$ which it should pass to gain a Lorentz factor $\gamma$:
		\begin{equation}
    		l_{{\rm acc}}=h\left(1-\sqrt{1-\frac{2\gamma} {\ell}} \right), \label{eq:acceleration}
		\end{equation}
	where $\ell=8\pi\eta B_{{\rm s}} eh_{\perp}^{2}\cos \left( \alpha\right)/ \left(Pc^{3}m \right)$.
 	Using the approximation $z_{1}+z_{2}\approx h$ we can write that
		\begin{equation}
    		\tilde{l}_{{\rm acc}}=\frac{\gamma m c^{3}P}{8\pi\eta B_{{\rm s}}e\cos\alpha}\frac{h}{h_{\perp}^{2}}.
        	\label{eq:acceleration_app}
		\end{equation}
	Note that for Lorentz factors that are considerably smaller than the maximum value, the discrepancy in the above formula is about a factor of two.
      
	\subsection{Gap height}
    
	By knowing the acceleration potential drop in PSG $\Delta V$ we can evaluate the gap height $h$, which actually depends on the details of the avalanche pair production in the gap.
	First, we need to determine which process, Curvature Radiation (CR) or Inverse Compton Scattering (ICS), is responsible for the $\gamma$-photon generation in the gap region.
	In order to identify the proper process we need the following parameters: $l_{{\rm acc}}$ - the distance which a particle should pass to gain the Lorentz factor $\gamma$, $l_{{\rm p}}$ - the mean length a particle (electron and/or positron) travels before a $\gamma$-photon is emitted, and $l_{{\rm ph}}$ - the mean free path of the $\gamma$-photon before being absorbed by the magnetic field.
   
		\subsubsection{Particle mean free path}
        
                   % ~/Programs/education/lines/lines/py (sets 555 and 556)
 		\begin{figure*}
    		\begin{center}
        		\includegraphics[width=7cm]{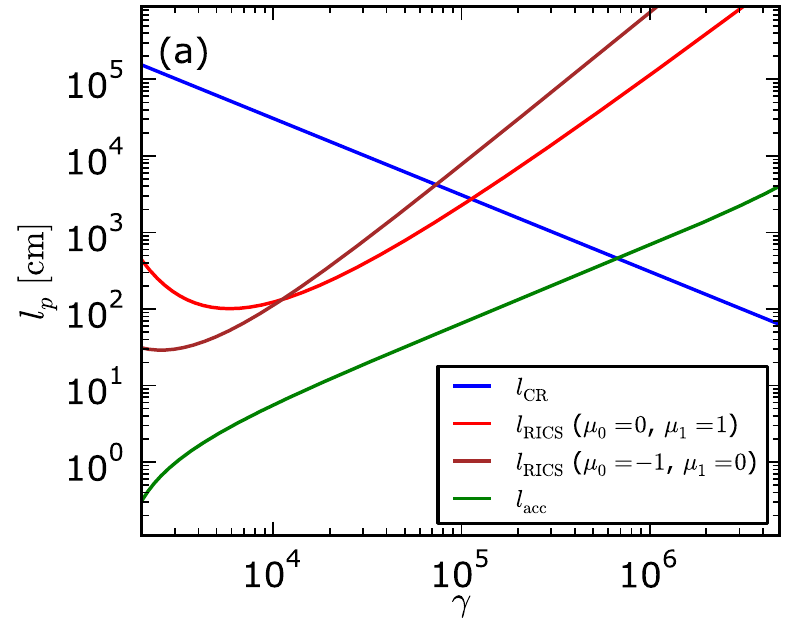}\includegraphics[width=7cm]{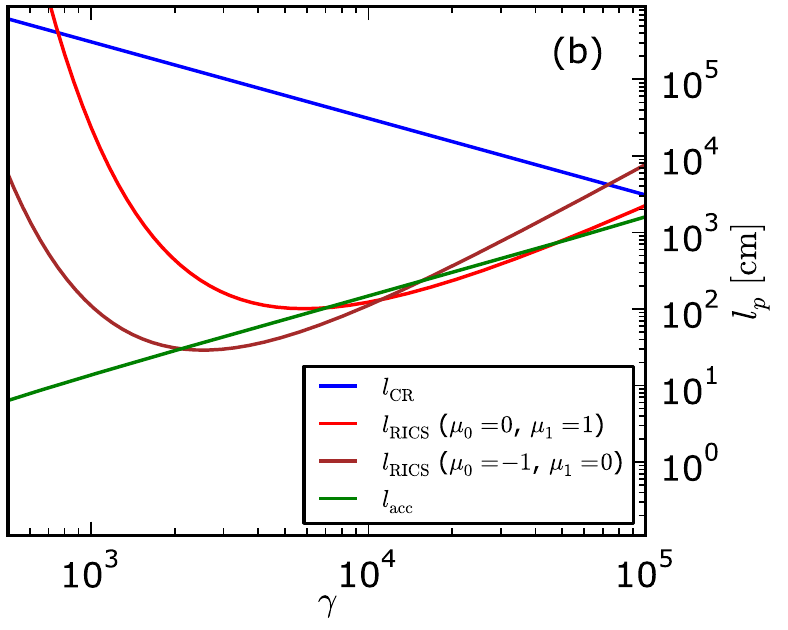}
			\end{center}
			\caption{Dependence of the mean free path of the primary particle on Lorentz factor $\gamma$ for both the CR and ICS processes.
	Panel (a) corresponds to calculations for a 'high-acceleration' potential drop ($h_{\perp}=3\,{\rm m}$ and $\eta=1$), while panel (b) corresponds to calculations for a 'low'-acceleration potential drop ($h_{\perp}=1\,{\rm m}$ and $\eta=0.1$). 
	The acceleration paths on both panels were calculated for the same pulsar parameters ($B_{14}=3.5$, $T_{6}=4.4$, $\Re_{6}=1$, $P=1$, $\alpha=10^{\circ}$).
%	Note that for the RICS process the particle mean free paths were calculated for optimal conditions (just above the polar cap). 
			}
			\label{fig:le_gamma}
		\end{figure*}
        
The mean free path of a particle (electron and/or positron) $l_{{\rm p}}$ can be defined as the mean length that a particle passes until a $\gamma$-photon is emitted. In the case of the CR particle, mean free path can be estimated as a distance that a particle with a Lorentz factor $\gamma$ travels during the time which is necessary to emit a curvature photon \citep[see][]{1997_Zhang_a}		\begin{equation}
    		l_{{\rm CR}}\sim c\left(\frac{P_{{\rm CR}}}{E_{\gamma,{\rm CR}}}\right)^{-1}=\frac{9}{4}\frac{\hbar\Re c}{\gamma e^{2}},\label{eq:le_cr}
		\end{equation}
where $P_{{\rm CR}}=2\gamma^{4}e^{2}c/3\Re^{2}$ is the power of CR, $E_{\gamma,{\rm CR}}=3\hbar\gamma^{3}c/2\Re$ is the photon characteristic energy, and $\Re$ is the curvature radius of the magnetic field lines.

For the ICS process calculation of the particle mean free path $l_{{\rm ICS}}$ is not as simple as that of the CR process. Although we can define $l_{{\rm ICS}}$ in the same way that we defined $l_{{\rm CR}}$, it is difficult to estimate the characteristic frequency of emitted photons. We have to take into account photons of various frequencies with various incident angles.
An estimation of the mean free path of an electron (or positron) to produce a photon is in \citet{1985_Xia}
		\begin{equation}
    		l_{{\rm ICS}}\sim\left[\int_{\mu_{0}}^{\mu_{1}}\int_{0}^{\infty}\sigma^{\prime}\left(\epsilon,\mu\right)\left(1-\beta\mu_{i}\right)n_{{\rm ph}}\left(\epsilon\right)d\epsilon d\mu\right]^{-1}.
            \label{eq:le_ics}
		\end{equation}
	Here $\epsilon$ is the incident photon energy in units of $mc^{2}$, $\mu=\cos\psi$ is the cosine of the photon incident angle, $\beta=v/c$ is the velocity in terms of speed of light, $\sigma^{\prime}$ is the cross section of ICS in the particle rest frame, 
		\begin{equation}
    		n_{{\rm ph}}\left(\epsilon,\, T\right)d\epsilon=\frac{4\pi}{\lambda_{c}^{3}} \frac{\epsilon^{2}}{\exp \left( \epsilon/\mho \right) -1}d\epsilon
            \label{eq:nph}
		\end{equation}
	represents the photon number density distribution of semi-isotropic blackbody radiation, $\mho=kT/mc^{2}$, $k$ is the Boltzmann constant, and $\lambda_{c}=h/mc=2.424\times10^{-10}$ cm is the electron Compton wavelength.
	A detailed description of how to calculate $\sigma^{\prime}$ can be found in \cite{2000_Gonthier} (see also \citealt{2013_Szary}).

	Since, both the photon density and incident angles ($\mu_{0}$ and $\mu_{1}$) change with increasing altitude, in our calculations we take into account both of those effects.
	Thus we replace $n_{\rm ph}\left(\epsilon,\, T\right)$, $\mu_0$ and $\mu_1$ with the photon density at location $\bf L$, $n_{\rm sp}\left(\epsilon,\, T,\, {\bf L}\right)$, the highest and lowest angle between the photons and particle at $\bf L$, $\mu_{\rm min} \left({\bf L} \right)$ and $\mu_{\rm max}\left( {\bf L} \right)$, respectively.
	We should expect two modes of ICS: resonant and thermal-peak.
	The Resonant ICS (RICS) takes place if the photon frequency in the particle rest frame is equal to the electron cyclotron frequency, $\epsilon^{\prime} = \epsilon \gamma (1-\beta \mu)=\epsilon_{B} $, here $\epsilon_B=eB/(mc)$.
	Thus, for RICS the mean free path of particles can be calculated as:
		\begin{equation}
    		l_{{\rm RICS}}\approx\left[\int_{\mu_{{\rm min}}\left(L\right)}^{\mu_{{\rm max}}\left(L\right)}\int_{\epsilon_{_{{\rm res}}}^{{\rm ^{min}}}}^{\epsilon_{_{{\rm res}}}^{{\rm ^{max}}}}\left(1-\beta\mu\right)\sigma^{\prime}\left(\epsilon,\mu\right)n_{{\rm sp}}\left(\epsilon,\, T,\, {\bf L}\right)d\epsilon d\mu\right]^{-1}, 
            \label{eq:ics_free_path}
		\end{equation}
	where the limits of integration over energy, $\epsilon_{{\rm _{res}}}^{{\rm ^{min}}}$ and $\epsilon_{{\rm _{res}}}^{^{{\rm max}}}$, are chosen to cover the resonant energy.
    In our calculations, we use such limits to include the region where the integrated function from its maximum decreases up to about two orders of magnitude:
		\begin{equation}
		\epsilon_{_{{\rm res}}}^{^{{\rm max/min}}}=\frac{\epsilon_{_{B}}\pm\frac{3}{2}\sqrt{11}\Gamma}{\gamma\left(1-\beta\mu\right)}.
		\end{equation}
	Here $\Gamma$ is the finite width introduced to describe the decay of an excited intermediate particle state, which depends on the strength of the magnetic field.  
	To describe the strength of the magnetic field we use $\beta_{q}=B/B_{q}$, where $B_{q}=m^{2}c^{3}/e\hbar=4.413\times10^{13}\,{\rm G}$ is the critical magnetic field strength.
    In the $\beta_{q}\ll 1$ regime, the cyclotron decay width assumes the well-known result $\Gamma\approx 4 \alpha_{f}\epsilon_{_{B}}^{2}/3$ in dimensionless units. 
    When $\beta_{q}\gg 1$, quantum and recoil effects generate $\Gamma\approx\alpha_{f}\epsilon_{_{B}}\left(1-1/\tilde{e}\right)$, here $\tilde{e}$ is the Euler's number \citep[see, e.g.,][]{2011_Baring}.

	The thermal-peak ICS (TICS) includes all scattering processes of photons with frequencies around the maximum of the thermal spectrum.
	In our calculations we adopt $\epsilon_{_{{\rm th}}}^{{\rm ^{min}}}\approx 0.05\epsilon_{{\rm _{th}}}$, and $\epsilon_{_{{\rm th}}}^{{\rm ^{{\rm max}}}}\approx2\epsilon_{_{{\rm th}}}$ as the limits of integration over energy in Equation \ref{eq:ics_free_path}. Here $\epsilon_{_{{\rm th}}}=2.82kT/\left(mc^{2}\right)$ is the energy, in units of $mc^{2}$, at which blackbody radiation with temperature $T$ has the largest photon number density.
    
	Figure \ref{fig:le_gamma} shows the dependence of particle mean free paths on the Lorentz factor $\gamma$ for a sample of pulsar parameters.
    Let us note that these free paths do not depend on the gap height $h$ (see Equations \ref{eq:acceleration}, \ref{eq:le_cr}, and \ref{eq:le_ics}).
	Although the presented results do not allow to define the gap height unambiguously, we can find which process is responsible for generation of $\gamma$-photons in PSG.
    Note that in order to produce a $\gamma$-photon, the particle mean free path has to be smaller that the acceleration path $l_{\rm p}<l_{\rm acc}$.
	Otherwise, the particles will be accelerated to higher energies before they start to emit photons (either by CR or ICS).
	The 'low-acceleration' gaps (e.g. due to narrow sparks - small $h_\perp$, or high ions density - high $\eta$) are dominated by ICS.  
    In such a case the $\gamma$-photon emission starts when particles reach Lorentz factor $\gamma_{\rm c}^{\rm ICS} \gtrsim 10^{3}$. 
	The 'high-acceleration' gaps (wide sparks - large $h_\perp$ or low ions density - small $\eta$), on the other hand, are dominated by CR.
    The characteristic Lorentz factor of primary particles in the CR-dominated gap is $\gamma_{\rm c}^{\rm CR} \gtrsim 10^6$.
       
        \begin{figure*}
            \begin{center}
                \includegraphics[width=14cm]{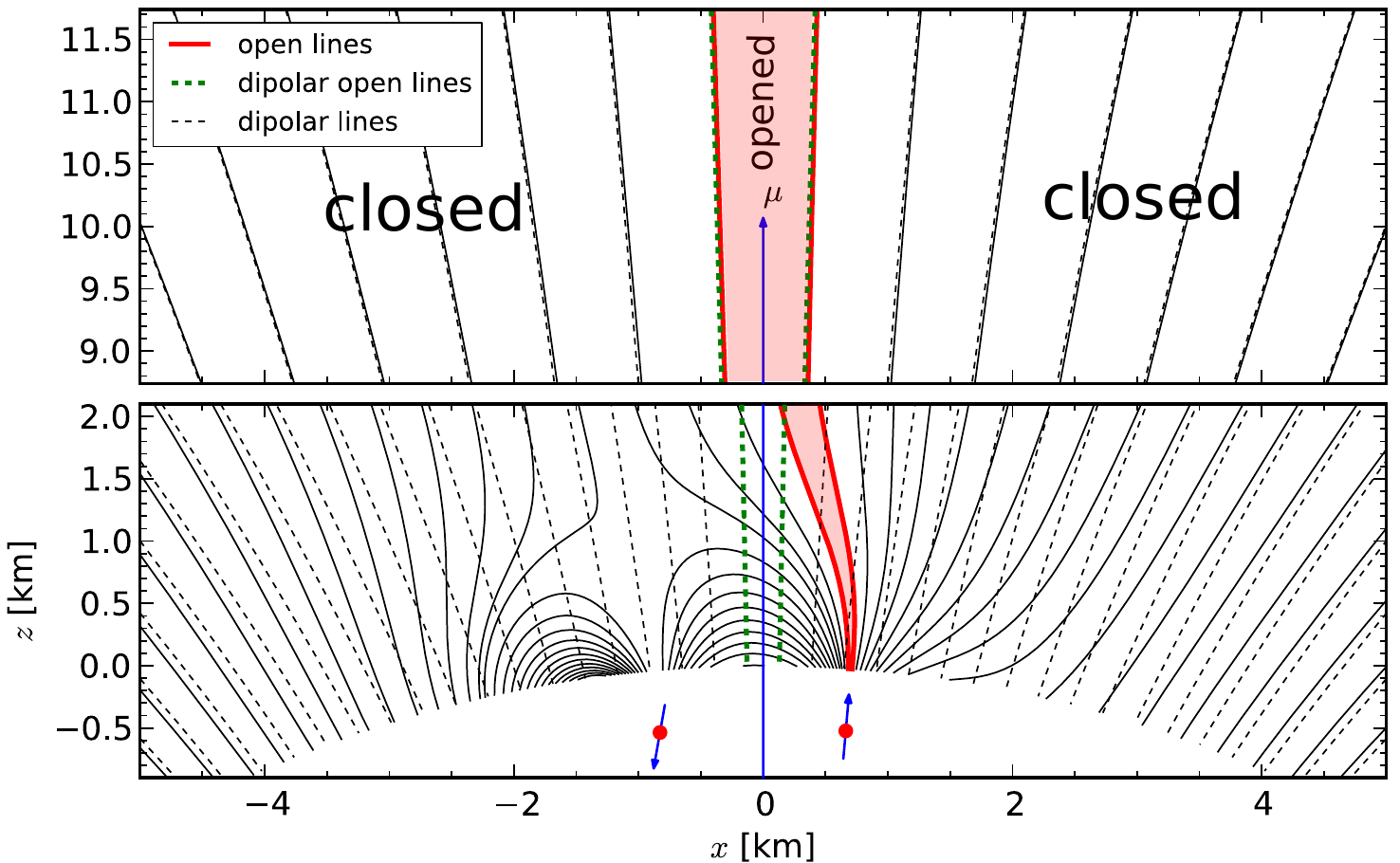}
            \end{center}
            \caption{Sample structure of the non-dipolar surface magnetic field. The structure was obtained using two crust anchored anomalies located at: ${\bf r_{1}}=\left(0.95R,2\,^{\circ},0\,^{\circ}\right)$ and ${\bf r_{1}}=\left(0.95R,5\,^{\circ},180\,^{\circ}\right)$ with the dipole moments ${\bf B_{\rm m1}} = \left(3\times10^{-3} B_{\rm d},\,15^{\circ},\,0^{\circ}\right)$, ${\bf B_{\rm m2}} = \left(3\times10^{-3} B_{\rm d},\,150^{\circ},\,0^{\circ}\right)$ respectively (blue arrows). The influence of the anomaly is negligible at distances $D\gtrsim3.2R$, where $B_{{\rm m}}/B_{{\rm d}}\approx m/d=3\times10^{-3}$.}
            \label{fig:magnetic_lines}
        \end{figure*}
       
		\subsubsection{Photon mean free path \label{sec:photon}}
        
	The photons with energy $E_{\gamma}>2mc^{2}$ propagating obliquely to the magnetic field lines can be absorbed by the field, and as a result, an electron-positron pair is created.
	
   The calculations of pair production attenuation coefficients done by \cite{2010_Medin} show that in strong magnetic fields ($\beta_q \gtrsim 0.2$) pairs are produced by photons almost immediately upon reaching the first threshold for pair production.
   Thus, if $\beta_q \gtrsim 0.2$ the created pairs will be in the low Landau levels ($n\lesssim 2$) and the mean free path can be estimated as
    
		\begin{equation}
			l_{{\rm ph}}\approx\Re\frac{2mc^{2}}{E_{\gamma}}, \label{eq:lph_strong}
		\end{equation}
	
    On the other hand, if $\beta_{q}\lesssim0.2$ the pairs are created at higher Landau levels, and an asymptotic approximation derived by \cite{1966_Erber} can be used to calculate the mean free path of a photon
		\begin{equation}
    		l_{{\rm ph}}=\frac{4.4}{(e^{2}/\hbar c)}\frac{\hbar}{mc}\frac{B_{q}}{B\sin\Psi}\exp\left(\frac{4}{3\chi}\right),
            \label{eq:lph_erber}
		\end{equation}
 
		\begin{equation}
    		\chi\equiv\frac{E_{\gamma}}{2mc^{2}}\frac{B\sin\Psi}{B_{q}}\hspace{1cm}(\chi\ll1),
		\end{equation}
	where $\Psi$ is the angle of intersection between the photon and the local magnetic field.
     
     A process which plays an important role in photon propagation, and thus pair multiplicity, is photon splitting. 
     For a high enough magnetic field photon may split before it produces an electron-positron pair 
     As was shown by \cite{2001_Baring, 2010_Medin}, if $\beta_q \gtrsim 0.5$ the $\perp$-polarised photons split before reaching the first threshold.
     In our calculations we use the kinetic selection rule in which only $\perp\rightarrow\parallel+\parallel$ process is allowed. 
     To calculate the optical depth, $\tau_{\rm sp}$ we adopt the numerical calculations of \cite{1997_Baring} \citep[see, e.g.][for more details]{2010_Medin,2013_Szary}. 
     Whenever $\tau_{\rm sp} \geq 1$ the $\perp$-polarised photon is turned into two $\parallel$-polarised photons.
     We assume that the energy of parent photon is equally distributed between both newly created photons travelling in the same direction as the parent photon.
    
        \subsubsection{Cascade simulations}
                
        In order to simulate progression of a primary particle in the acceleration region, we assume non-dipolar structure of surface magnetic field.
        The non-dipolar configuration was obtained using the model proposed by \cite{2002_Gil} in which the actual surface magnetic field results from the superposition of a global dipole magnetic field $\mathbf{B_{\rm d}}$ and crust anchored small scale magnetic anomalies $\mathbf{B_{\rm m}}$.
        In Figure \ref{fig:magnetic_lines} we present a sample configuration of the surface magnetic field used in our calculations. 
        The dashed lines correspond to the dipolar configuration of the magnetic field lines and by green colour we show the last open magnetic field lines. 
        The solid lines correspond to an actual configuration of the magnetic field and by the red colour we show an actual open field line region.
        Note that our approach allows to reproduce two observational facts: (1) a very small size of the polar cap (e.g., derived from X-ray observations) and (2) the dipolar configuration of magnetic field at altitudes where the radio emission is produced.
        In fact, already at the distance $D \approx R$, the magnetic field structure is dominated by the global dipole component (see the top panel in Figure \ref{fig:magnetic_lines}).
        The presented configuration was calculated for $P=1.24$, $\dot{P}=7.1\times 10^{-15}$ ($B_{\rm d}=6 \times 10^{12}\,{\rm G}$), $\alpha=70^{\circ}$  and results in the following parameters of the magnetic field at the polar cap region: $B_{\rm s}=2.2\times 10^{14}\,{\rm G}$, $\Re_{6}=0.6$.
        In order to obtain, for instance, different values of the surface magnetic field and the curvature radius we modify the location and magnetic moment of the crust-anchored anomalies. 
        
        %In this paper we use the approach of calculating the pair cascades developed by \cite{2010_Medin} which has been applied to cases with non-dipolar structure of magnetic field.
        To track progression of a primary particle through the acceleration region of a given pulsar ($P$, $\dot{P}$, $\alpha$) with assumed structure of the surface magnetic field ($B_{\rm s} \rightarrow T_{\rm s}$, $\Re$) we perform the following steps:
        (1) Depending on the mode in which the gap operates, we establish the required gap parameters: the gap height $h$, the spark half-width $h_{\perp}$, and shielding factor $\eta$ (see Sections \ref{sec:psg-off} and \ref{sec:psg-on}).
        (2) Then, a positron is accelerated in a stepwise fashion from the neutron star surface up to the gap height. 
        The particle travels in a sample spark located along the central open field line with the length of each step fixed to $\Delta s = 10^{-3} h$.
        Note that the Lorentz factor of the primary particle is proportional to $(h-z_1)^2$, where $z_1$ is the starting altitude (see Equation \ref{eq:gamma_z}). 
        Thus, assuming for instance, $z_1=0.3h$ the maximum Lorentz factor of the primary particle $\gamma \approx 0.5 \gamma_{\rm max}$ which does not change the results significantly.
        (3) For each step we calculate the particle mean free path for CR or ICS depending on the gap mode. The particle radiates a photon when $\tau_{\rm rad} = \sum l_{\rm p,i} / \Delta s \geq 1$. Depending on the emission process we randomly assign the photon a polarisation. For CR we assign one $\perp$ to every seven $\parallel$ photons resulting in $75\%$ averaged parallel polarisation, while for ICS we assign polarisation in the radio of one $\perp$ to every $\parallel$-polarised photon ($\beta_q > 1$, see. e.g. \cite{2010_Medin}).
        The energy of a primary particle is reduced by the energy of an emitted photon and continues propagation through the acceleration region.
        
        Each emitted photon is tracked in a similar stepwise fashion from the emission point ($\Delta s_{\rm ph} = 10^{-3} h$).
        For every step we estimate the photon mean free path to find the height at which a photon is absorbed by the magnetic field and an electron-positron pair is produced.
        The pair is produced when the optical depth for pair production $\tau_{\rm pr} = \sum l_{\rm ph,i} / \Delta s_{\rm ph} \geq 1$ .
        At the same time, for every step, we calculate the optical depth for $\perp$-polarised photon splitting. 
        A photon splits when $\tau_{\rm sp} = \sum R^{\rm sp}_{\perp \rightarrow \parallel \parallel} \geq 1$, here $R_{\rm sp, i}^{\perp \rightarrow \parallel \, \, \parallel}$ is the attenuation coefficient for photon splitting \citep[see][for more details]{1997_Baring, 2001_Baring, 2010_Medin, 2013_Szary}.
        At the point of photon splitting two new photons are created with the energy of the parent photon evenly divided among them.
        The newly created photons travel in the same direction as the parent photon.
        
		\subsubsection{Results}
        
\begin{figure}
    		\begin{centering}
        		\includegraphics[width=8cm]{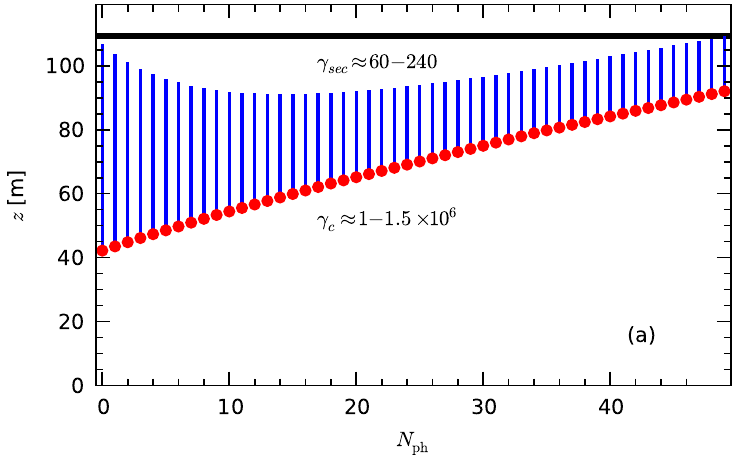}
        		\includegraphics[width=8cm]{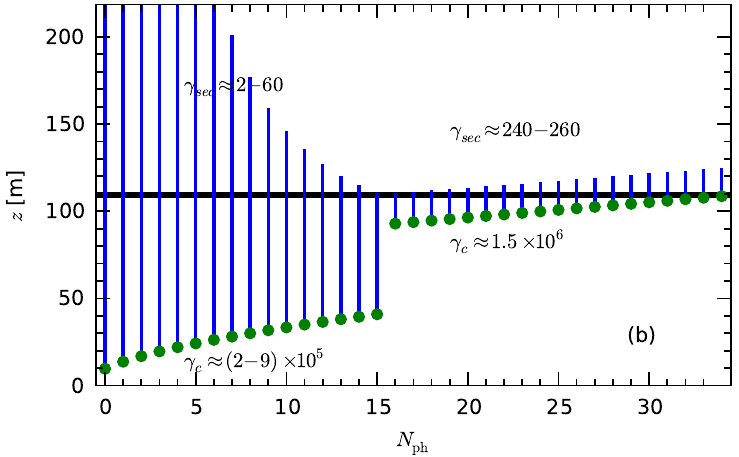}
    		\end{centering}
    		\caption{Cascade formation in a CR-dominated gap. Blue lines represent the free paths of $\gamma$-photons, while the filled circles correspond to places of photon emission.
               	Panel (a) includes the free paths of photons which produce pairs below ZPF while panel (b) includes the free paths of photons which produce pairs above ZPF.
				The results were obtained using the following parameters: $N_{\rm ph}^{\rm CR}=50$, $B_{\rm s}=2.0 \times 10^{14}\,{\rm G}$, $B_{\rm d}=2.6 \times 10^{12}\,{\rm G}$, $T_{s}=3.3 \, {\rm MK}$, $P=0.58 \, {\rm s}$, $\Re_{6}=0.4$, and $\alpha=36^{\circ}$.
    			\label{fig:cr_sol}}
		\end{figure}

	Figure \ref{fig:cr_sol} shows progression of a primary particle and photon mean free paths of $\gamma$-photons produced in the CR-dominated gap.
	The gap height was chosen to allow production of $N_{\rm ph}^{\rm CR}=50$ photons within the acceleration region. 
	The primary particle should travel a distance comparable with a gap height $l_{\rm acc}\approx h/2$ in order to gain an energy required to initiate efficient production of photons ($\gamma_{\rm c}^{\rm CR} \gtrsim 10^6$), and thus production of secondary particles (see Figures \ref{fig:le_gamma} and \ref{fig:cr_sol}).
	Most of the CR photons produce pairs at about the same height, $z\approx h$, in the region where the acceleration potential is almost equal to zero, hereinafter we will call this region the Zero-Potential Front (ZPF).
	The newly created particles have much lower Lorentz factors as compared with the primary particle ($\gamma \approx 10^2$), and thus for emission of CR photons only electrons are important.
	Note that the newly created electrons should be first decelerated and then accelerated towards the stellar surface (travel a distance $l_{\rm acc}\approx h/2$) in order to start efficient production of CR photons.
    % radiation/gap.py find_solution_cr_psgoff_plot(), plot_solution_cr(), plot_solution_acr()

\begin{figure}
    		\begin{center}
        		\includegraphics[width=8cm]{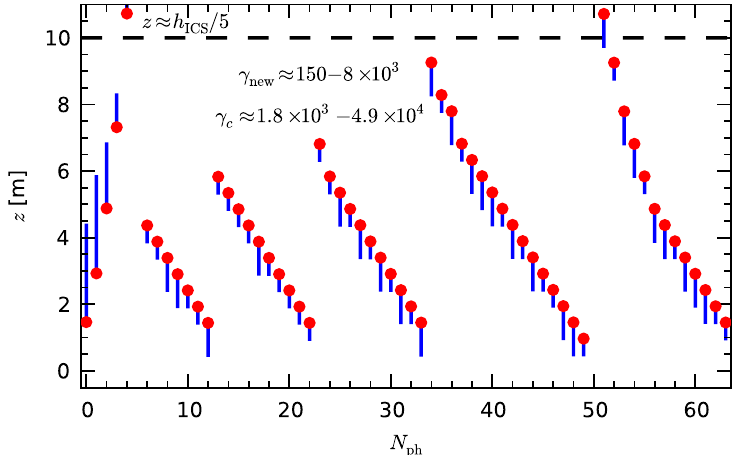}
    		\end{center}
    		\caption{Cascade formation in an ICS-dominated gap. Blue lines represent the free paths of $\gamma$-photons, while the filled circles correspond to places of photon emission.
    		The results were obtained using the following parameters: $h_{\perp}=1\,{\rm m}$, $N_{\rm ph}^{\rm ICS}=64$, $B_{\rm s}=2.1 \times 10^{14}\,{\rm G}$, $B_{\rm d}=2.6\times10^{12}\,{\rm G}$, $T_{\rm s}=3.5 \,{\rm MK}$, $P=0.58\,{\rm s}$, $\Re_{6}=0.3$, and $\alpha=36^{\circ}$.
    		\label{fig:ics_sol}}
		\end{figure}

	On the other hand, primary particles in the ICS-dominated gap reach a characteristic value $\gamma_{\rm c}^{\rm ICS}\approx 10^4$ at altitudes that are considerably smaller than the gap height $l_{\rm acc}\ll h$ (see Figure \ref{fig:le_gamma}).
	Figure \ref{fig:ics_sol} shows schematically the locations at which $\gamma$-photons are emitted by ICS. 
    The first $\gamma$-photon is produced already at altitudes of about a few metres and then well below ZPF is converted to an electron-positron pair.
	The energy of $\gamma$-photons produced by RICS depends on the Lorentz factor of the primary particles and on the magnetic field strength.
	RICS photons upscattered in an ultrastrong ($\beta_q>1$) magnetic field gain a significant part of the energy of the scattering (primary) particle.
    The newly created electron-positron pairs have energies comparable with the energy of primary particle ($\sim \gamma_{\rm c}^{\rm ICS}/2$) and enhance gamma-photon emission in the lower part of the gap.
	Furthermore, RICS in ultrastrong magnetic fields produces approximately the same amount of photons with $\parallel$ and $\perp$ polarisation \citep{2000_Gonthier}, while most of the photons produced by CR are $\parallel$-polarised ( between $50\%$ and $100\%$ depending on photon energy).
    Thus, the photon splitting of $\perp$-polarised photons further enhance number of $\gamma$-photons and, therefore, number of pairs in the lower part of the gap (see Section \ref{sec:photon} and references therein).
	% gap_ics.py: m_end=180, read_data(505), get_start_point(0, 0, midline=True), find_solution_hperp(h_perp=100.), plot_solution()

	\section{Possible scenarios of the gap breakdown} \label{sec:results}
        
	The differences between the CR- and ICS- dominated gaps that we mention above have crucial consequences on the cascade formation process.
    In general an acceleration gap can break according to two scenarios: (1) production of dense enough plasma with charge density $\rho_+ = \rho_{\rm GJ}$, or (2) overheating of the stellar surface to temperature $T_{s}>T_{\rm crit}$ and thus extraction of ions with net charge density $\rho_i \geq \rho_{\rm GJ}$.
	Cooling time of the polar cap is very short ($\tau_{{\rm cool}}\lesssim 10^{-8}\,{\rm s}$, see \cite{2003_Gil}), thus, to sustain its temperature just below the critical value a continuous backflow of relativistic particles is required.

    	\subsection{PSG-off mode}
   
   	Curvature emission by a primary particle is effective for Lorentz factors $\gamma>10^5$ (when $l_{\rm CR} \leq l_{\rm acc}$). 
    An equilibrium between acceleration and deceleration (by reaction force) would be established if the CR power were equal to the ”electric power”. 
    For a typical values ($\Re_6 \approx 1$, $\gamma_{\rm c}\approx 10^6$), the reaction force is not high enough to stop acceleration by the electric field.
	As we have shown, in the case of CR-dominated gap pairs are created in the upper part of the gap.
    The time which has to pass from the beginning of primary particle acceleration to heating of the polar cap by newly created particles is too long ($\sim 2h/c \approx 6 \times 10^{-7}\,{\rm s}$ - for a typical gap height $h=100\,{\rm m}$) to keep the surface temperature close to the critical value.
	Therefore, while the CR-dominated gap is open the surface temperature $T_{\rm s}<T_{\rm crit}$ and the screening factor $\eta \approx 1$, hence we call this mode the PSG-off mode.
	The primary particles in the PSG-off mode are very energetic $\gamma\approx10^{6}$, and hence the density of backstreaming particles required to heat the polar cap to the critical temperature, $\rho_{c}$, is much lower than the Goldreich-Julian co-rotational density.
%    The growth of particle density will continue to the point when the backstreaming particles heat up the surface to a temperature equal to or higher than the critical temperature.
	To describe this difference we use the overheating parameter $\kappa=\rho_{c}/\rho_{\rm GJ}$.
	On the one hand, the flux of backstreaming particles with the charge density $\kappa n_{\rm GJ}$ can be estimated as $\kappa n_{\rm GJ} e \Delta V c$, here $n_{\rm GJ} = \rho_{\rm GJ}/e$ is the Goldreich-Julian co-rotational number density.
	On the other hand, we can estimate the flux required to heat up the polar cap to the critical temperature as $\sigma T_{\rm crit}^{4}$, here $\sigma$ is the Stefan-Boltzmann constant.
	Finally, using the relation $\Delta V=\gamma_{\rm max} mc^{2}/e$ we can estimate the overheating parameter as follows:
	\begin{equation}
    		\kappa=\frac{\sigma\, T_{\rm crit}^{4}}{n_{{\rm GJ}}\,\gamma_{\rm max}\, mc^{3}}.
    		\label{eq:overheating_parameter}
		\end{equation}

		\subsection{PSG-on mode}
        
	In the case of the ICS-dominated gap pairs are created just above the stellar surface providing constant backflow of relativistic particles required for heating of the polar cap.
	The heating of the surface will sustain the outflow of iron ions from the crust, maintaining $\eta<1$, hence we call this mode the PSG-on mode.
 	The actual potential drop $\Delta V$ should be thermostatically regulated and a quasi-equilibrium state should be established in which heating due to the electron/positron bombardment is balanced by cooling due to thermal radiation.
	The necessary condition for this quasi-equilibrium state is 
		\begin{equation}
    		\sigma T_{{\rm s}}^{4}=\eta e\Delta Vcn_{\rm GJ}. \label{eq:heating_condition}
		\end{equation}
	Here, we assume that the number density of relativistic backstreaming electrons is $\eta n_{{\rm GJ}}$.
	The Goldreich-Julian co-rotational number density can be expressed in terms of $B_{14}$ as $n_{{\rm GJ}}=6.94\times10^{12}B_{14}P^{-1}$.
	Using Equations \ref{eq:t_crit} and \ref{eq:heating_condition} we can find the acceleration potential drop that satisfies the heating condition:
		\begin{equation}
    		%\Delta V=7.3\times10^{5}\frac{\left(B_{14}^{1.1}+0.3\right)^{4}P}{\eta B_{14}}.
    		\Delta V=9.1 \times 10^{6} B_{14}^{2} P / \eta. \label{eq:potential_heating}
		\end{equation}
	The above equation may suggest that the acceleration potential drop is inversely proportional to the screening factor. In fact it is just the opposite.
	On the one hand, the ions presence in the gap lowers the potential drop (see Equation \ref{eq:potential_drop}), but on the other hand it also lowers the density of backstreaming particles.
    Thus, gaps with high ions density (small $\eta$) require higher acceleration potential in order to heat the polar cap to the critical temperature (see Equation \ref{eq:potential_heating}). 
	Knowing that $\Delta V=\gamma_{\rm max} mc^{2}/e$, we can calculate the maximum Lorentz factor of the primary particles in ththee  PSG-on mode $\gamma_{\rm max} = 5.3 \times 10^{3} B_{14}^{2} P / \eta $.
	Using Equations \ref{eq:potential_drop} and \ref{eq:potential_heating} for specific pulsar parameters we can define a product of the two main parameters of PSG, namely the screening factor $\eta$ and the spark half-width $h_{\perp}$:
		\begin{equation}
			\eta h_{\perp} = 15 P B_{14}^{0.5} \left | \cos \alpha \right |^{-0.5} \label{eq:eta_hperp}
		\end{equation}
As in the PSG-on mode the temperature of the polar cap is in quasi equilibrium with the backstreaming particles (see Figure \ref{fig:thermostat}) the gap can break only due to production of a dense enough plasma \mbox{$n_{p} \gg \eta n_{{\rm GJ}}$} in the gap region.
	Hence, the multiplicity in the PSG-on mode is much higher than the multiplicity of CR-dominated gaps. % not a multiplicity in CR case

        \subsection{Finding gap parameters \label{sec:finding_height}}
        
                \begin{figure}
                        \begin{center}
                                \includegraphics[width=8cm]{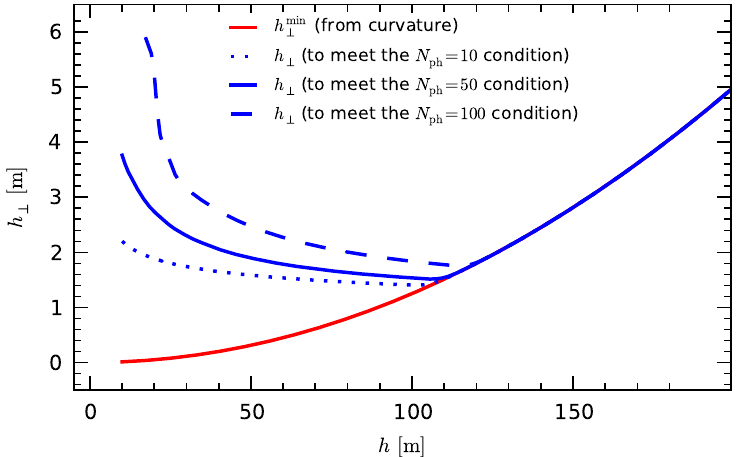}
                        \end{center}
                        \caption{Dependence of a spark half-width on the gap height in the PSG-off mode. 
            The results were obtained using the following pulsar parameters: $B_{14}=2.0$, $T_{6}=3.3$, $P=0.58\,{\rm s}$, $\Re_{6}=0.3$, $\alpha = 36^{\circ}$. \label{fig:hperp_h_nph} }
                \end{figure}        
        
\begin{figure*}
			\begin{center}
				\includegraphics[width=15.5cm]{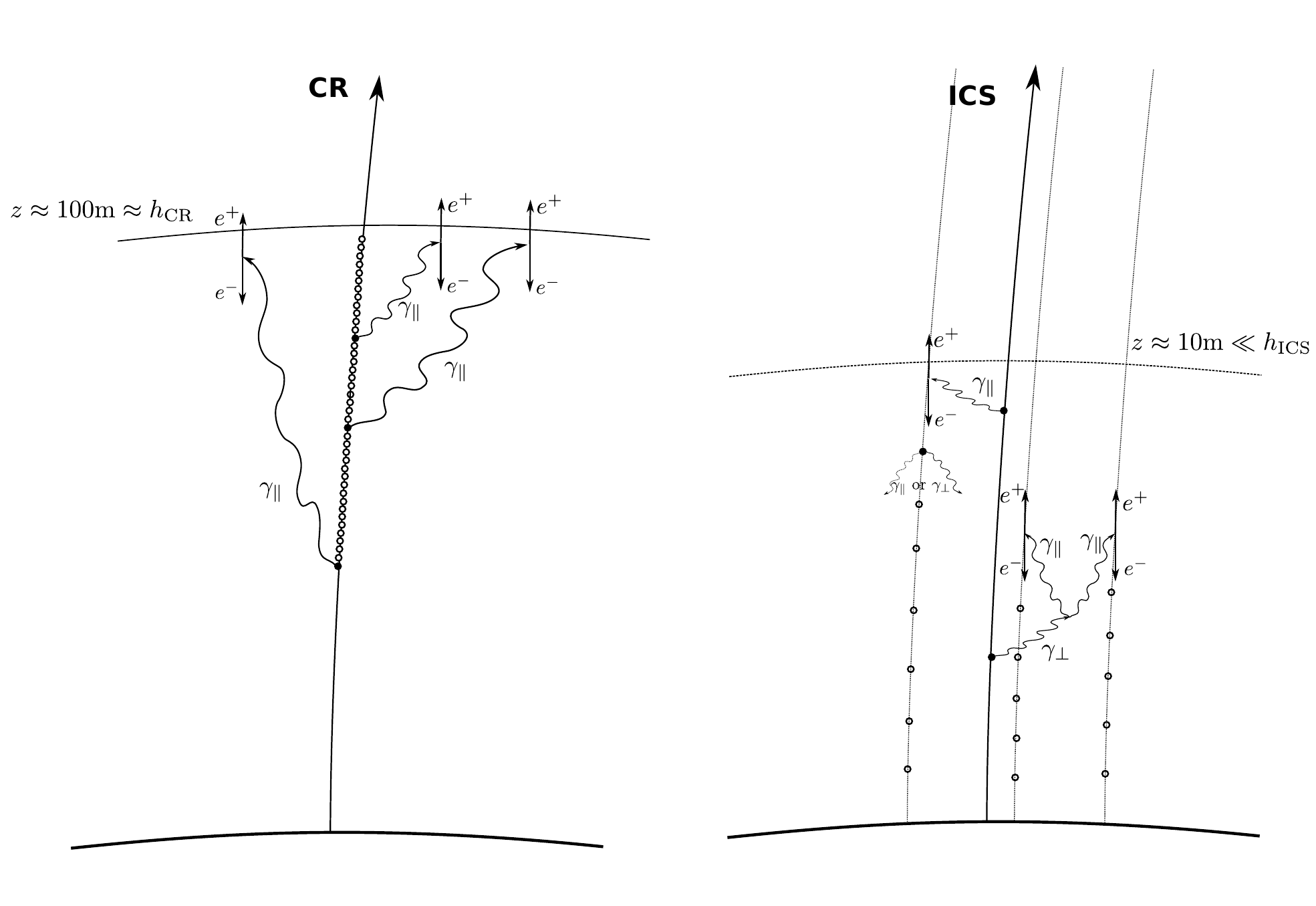}
			\end{center}
			\caption{Sketch of differences in a cascade formation for the CR-dominated gap (left panel) and the ICS-dominated gap (right panel). 
    		In order to increase readability, only a few points (filled circles) are shown which correspond to altitudes where $\gamma$-photons are emitted.
			The unfilled circles correspond to places where $\gamma$-photons are also emitted, but those photons (and their progression) are not included in the diagram. 
            Note that for the ICS-dominated gap we plot only the bottom (active) part of the gap ($z\ll h_{{\rm ICS}}$), furthermore, points of radiation are tracked only for the first population of newly created particles. 
            The avalanche nature of the ICS-dominated gap will result in a much higher multiplicity and continuous backflow of relativistic particles.
            \label{fig:schematic_ics_cr}}            
        \end{figure*}

        Figure \ref{fig:schematic_ics_cr} presents a sketch of a cascade formation for CR- and ICS-dominated gaps.
        Since in different modes the breakdown of the gap is done according to two completely different scenarios we use separate procedures to estimate gap parameters.
        
        \subsubsection{in PSG-off mode} \label{sec:psg-off}
In the PSG-off mode the spark region is free from ions $\eta \approx 1$, thus the heating condition (Equation \ref{eq:heating_condition}) is no longer satisfied.
	In order to close the CR-dominated gap surface below the spark has to be heated to temperature exceeding the critical value.
    To find the gap parameters in this mode we assume that a gap height and a spark half-width have to be connected in such a way that generated pairs can heat up the whole spark surface.
    Thus, we can write that for a given curvature radius this relation should be as follows:
		\begin{equation}
			h_{\perp}^{\rm max}=\Re - \sqrt{\Re^{2} - h^{2}}. \label{eq:psg.hperp_min}
		\end{equation} 
     On the other hand, we can estimate the acceleration potential $\Delta V$ (and thus the spark half-width $h_{\perp}^{N_{\rm ph}}$) required to produce a specified number of photons $N_{\rm ph}^{\rm CR}$ within a gap.
	Figure \ref{fig:hperp_h_nph} presents the dependence of both $h_{\perp}^{\rm min}$ and $h_{\perp}^{N_{\rm ph}}$ on the gap height.
	In order to find the gap height, we assume $h_{\perp}^{\rm min}=h_{\perp}^{N_{\rm ph}}$, which results in a gap that allows both overheating of the entire spark surface by backstreaming particles and the creation of the required number of photons $N_{\rm ph}^{\rm CR}$ within a gap.  
	As results from the Figure, the gap height in the PSG-off mode does not change significantly with $N_{\rm ph}^{\rm CR}$, and
for these specific parameters of a pulsar it is $h \approx 120\,{\rm m}$.
	For historical reasons, hereafter unless stated otherwise, we use $N_{\rm ph}^{\rm CR}=50$ to calculate gap parameters in the PSG-off mode.
      
	In our calculations we use the algorithm presented in the left panel of Figure \ref{fig:flowchart} to find the gap height in the PSG-off mode for given pulsar parameters: a pulsar period $P$, a surface magnetic field strength $B_{{\rm s}}$, and a curvature radius of field lines $\Re$, and a pulsar inclination angle $\alpha$.

        \subsubsection{in PSG-on mode} \label{sec:psg-on}
\begin{figure*}
			\begin{center}
				\includegraphics[width=15cm]{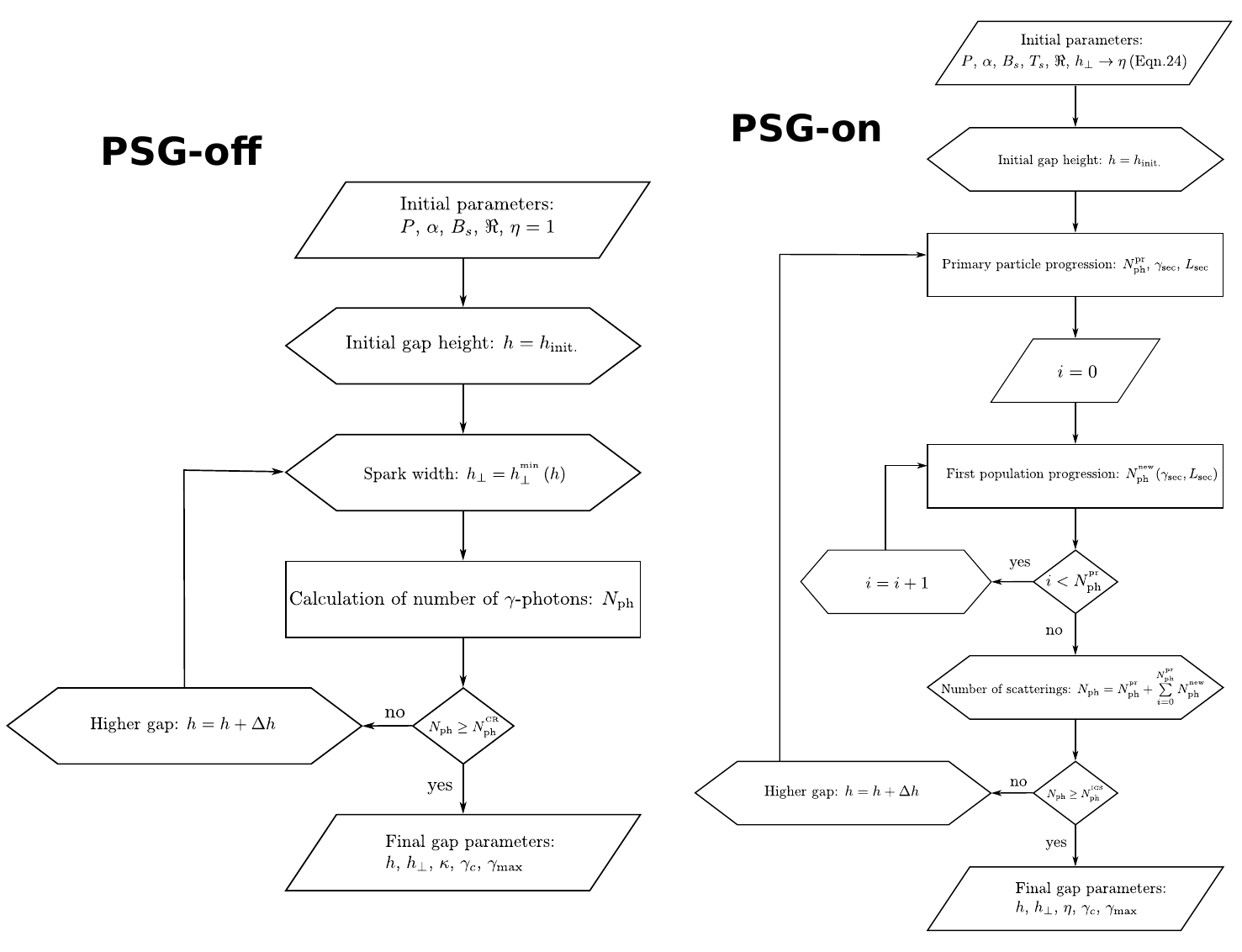}
			\end{center}
			\caption{Flowchart of the algorithm used to estimate the gap height in the PSG-off ({\it left panel}) and PSG-on modes ({\it right panel}).
                        The initial gap height from which we begin our calculations is an arbitrary set to $h_{{\rm init.}}=10\,{\rm m}$, while the step $\Delta h$ depends on the required accuracy. 
			The number of $\gamma$-ray photons created in a spark by a single primary particle is set to $N_{\rm ph}^{\rm CR}=50$ and $N_{\rm ph}^{\rm ICS}=10$ for PSG-off and PSG-on modes, respectively. \label{fig:flowchart}}
                \end{figure*}

	In the PSG-on mode, radiation of the surface just below the spark is in quasi-equilibrium with the flux of backstreaming particles.
    This quasi-equilibrium state prevents the gap
breakdown due to surface overheating, and gap can break only due to production of dense plasma.
	To describe the acceleration potential drop in the PSG-on mode we need information on screening factor $\eta$ (and thereby ions density) or on spark half-width $h_{\perp}$ (see Equations \ref{eq:potential_drop} and \ref{eq:eta_hperp}).
    Although such information cannot be directly determined from basic pulsar parameters ($P$, $B_{\rm s}$, $\Re$, $\alpha$), we can use X-ray and drift information to put constrains on the spark half-width \citep{2013_Szary}.
    For the observed sample of pulsars the spark half-width is in the range of $0.5-4\,{\rm m}$.
    Thus, in order to find the approximate gap parameters in the PSG-on mode we use $h_{\perp} = 2 \, {\rm m}$.
	The right panel in Figure \ref{fig:flowchart} presents the procedure of finding the gap height in the PSG-on mode.
    First we use Equation \ref{eq:eta_hperp} to estimate the screening factor $\eta$ which determines the electric field, and thus the particle acceleration.
    Then, for the initial gap height, we estimate the number of scatterings for a single outflowing primary particle $N_{\rm ph}^{\rm pr}$.
    The initial gap height from which we begin the height search is arbitrary set to $h_{\rm init.}=10\,{\rm m}$.
	We track the propagation of $\gamma$-photons produced by ICS on the primary particle to find the locations where pairs are created $L_i$. 
    Then we calculate propagation of newly created pairs through the acceleration region and we estimate the number of scatterings by every particle of the first population $N_{\rm ph}^{\rm i}$.
    If the total number of scatterings by the first population (including the primary particle) is $N_{\rm ph}<N_{\rm ph}^{\rm ICS}$ , we resume our calculations assuming a higher gap until the $N_{\rm ph}\geq N_{\rm ph}^{\rm ICS}$ is met.
	We use $N_{\rm ph}^{\rm ICS}=10$ to calculate gap parameters in the PSG-on mode.
    Note, however, that since in our estimate we include only particles from the first population (created by $\gamma$-photons emitted by a single primary particle) an actual number of created particles is much higher, thus higher multiplicity in the ICS-dominated gap.

        \subsection{Influence of the magnetic field}
        
	The conditions in PSG are mainly defined by the surface magnetic field.
	In Figure \ref{fig:h_b14}, panel (a) we present the dependence of the gap height on the surface magnetic field calculated according to the approach described in Section \ref{sec:finding_height}.
	It is clear that in the PSG-off mode the gap height decreases as the surface magnetic field increases.
    Furthermore, it also entails decrease of the spark half-width from $h_{\perp} = 2.6 \,{\rm m}$ for $B_{14}=0.7$ down to  $h_{\perp} = 1.0 \,{\rm m}$ for $B_{14}=5.0$.
    In the PSG-off mode for stronger magnetic fields the gap breaks before it has time to grow both in height and width.
    In the PSG-on mode, on the other hand, the gap height shows a minimum at a specific value of the magnetic field strength (for a given pulsar parameters it is $B_{14}\approx 4.$).
	This behaviour is the result of an increasing acceleration potential with an increasing surface magnetic field. 
    When the magnetic field strength exceeds the optimal value, which corresponds to acceleration when ICS is the most effective, the increase in the acceleration potential results in less effective scattering. 
    Panel (b) shows the dependence of the screening factor (or the overheating parameter $\kappa$ in
the PSG-off mode) on the surface magnetic field. 
	We can see that for stronger magnetic fields both $\eta$ and $\kappa$ increase, which means that: (1) the density of heavy ions above the polar cap in the PSG-on mode decreases, (2) the density of particles required to overheat the polar cap increases. 
    Let us note that the surface temperature $T_{\rm s}$ stays very near to the critical temperature $T_{\rm crit}$, which is shown on the top axis of the Figures.
    In panel (c), the red solid and dotted lines correspond to characteristic and maximum Lorentz factors ($\gamma_{\rm c}$, $\gamma_{\rm max}$) in the PSG-on mode, while the blue, dashed and dashed-dotted lines correspond to $\gamma_{\rm c}$ and $\gamma_{\rm max}$ in the PSG-off mode. 
    We see that especially in the PSG-off mode $\gamma_{\rm c}$ does not depend on the magnetic field strength.
    Note that in the PSG-off mode (CR-dominated gap), the characteristic Lorentz factor (the Lorentz factor for which most of the gamma photons are produced) slightly differs from the maximum value, $\gamma_{\rm c}\approx\gamma_{{\rm max}}$.
	In the PSG-on mode, on the other hand, $\gamma_{\rm c}\ll \gamma_{\rm max}$, which reflects the fact that most of the scatterings take place in the bottom part of the gap.
	Note that since the surface magnetic field and temperature are related (see Equation \ref{eq:t_crit}), there is a minimum value of the surface magnetic field  below which the gap breakdown in the PSG-on mode is not possible (for a given pulsar parameters it is $B_{{\rm min}}\approx 0.7\times 10^{14}\,{\rm G}$).
	Below $B_{\rm min}$ the flux of X-ray background photons is too low to enable efficient ICS.

        \subsection{Influence of the curvature radius}
        
    The curvature radius of the magnetic field lines affects the photon mean free path, namely the magnetic field with smaller radius of curvature (higher curvature) absorbs photons faster.
    In the case of the CR-dominated gap, the curvature radius influences also the energy of photons generated in the gap region, and hence the particle mean free path. 
	Therefore, the gap height in the PSG-off mode highly depends on the radius of curvature of the magnetic field lines (see Figure \ref{fig:h_re6}, panel a).
    In contrast, the gap height in the PSG-on mode is only slightly affected by changes in the curvature radius.
    In the case of the ICS-dominated gap the most important parameter which determines the gap height is the primary particle mean free path which does not depend on the curvature radius (see Equation \ref{eq:ics_free_path}). 
	The overheating parameter in the PSG-off mode inversely depends on the radius of curvature of the magnetic field lines (see Figure \ref{fig:h_re6}, panel b). 
    The smaller the curvature radius, the higher the overheating parameter, which corresponds to  narrower sparks.
    This is consistent with the expectation that for a smaller curvature radius of the field
lines, the gap breakdown is easier to develop and takes place before the sparks manage to grow in width.
	On the other hand, the screening factor in the PSG-off mode does not depend on the curvature of the magnetic field lines.
	With an increasing radius of curvature the Lorentz factor of primary particles (both $\gamma_{{\rm c}}$ and $\gamma_{{\rm max}}$) required to close the gap in the PSG-off mode also increases. 
    This reflects the fact that in order to produce a sufficient number of photons in the gap region, the primary particles should be accelerated to higher energies.
    The higher energies of the primary particles will increase the emitted $\gamma$-photon energy, thereby they will partly inhibit the growth of the photon mean free path.
    As mentioned above, the gap height in the PSG-on mode very weakly depends on the photon mean free path, thus both $\gamma_{{\rm c}}$ and $\gamma_{{\rm max}}$ are not affected by the increase in the radius of curvature.

        \subsection{Influence of the pulsar period}
	
    As we can see from Figure \ref{fig:h_p}a, the gap height in both modes highly depends on the pulsar period.
     Longer periods entail also an increase in the screening factor (in the PSG-on mode, see Equation \ref{eq:eta_hperp}) and in the overheating parameter (in the PSG-off mode).
     The increase of the overheating parameter in the PSG-off mode is caused by the decrease of the potential drop, and thereby the decrease of the maximum Lorentz factor of primary particles (see Equations \ref{eq:potential_drop}, \ref{eq:overheating_parameter} and Figure \ref{fig:h_p}c). 
     In the PSG-on mode, on the other hand, the acceleration potential does not depend on $P$ (see Equations \ref{eq:potential_heating} and \ref{eq:eta_hperp}).
     However, the efficiency of ICS process highly depends on the density of the X-ray background photons.
     Taking into account the that the polar cap is the main source of the background photons and that its area decreases with increasing pulsar period ($A_{\rm bb} \propto A_{\rm dp} \propto P^{-1}$), the ICS process becomes less and less efficient for pulsars with increasing periods.
 	Note that for periods longer than some specific value (for a given pulsar parameters it is $P_{{\rm max}}\approx5\,{\rm s}$), the screening factor in the PSG-on mode would exceed unity.
    This means that the PSG-on mode cannot be responsible for the gap breakdown for pulsars with such a long periods.
           \\
       
\begin{figure*}
			\centering{} \includegraphics{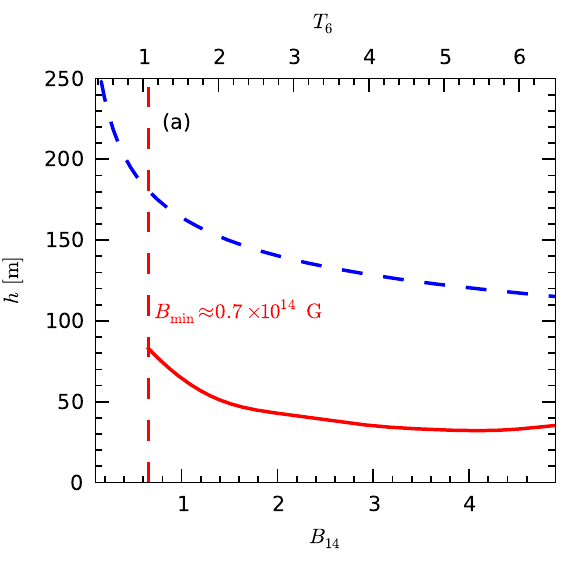} \includegraphics{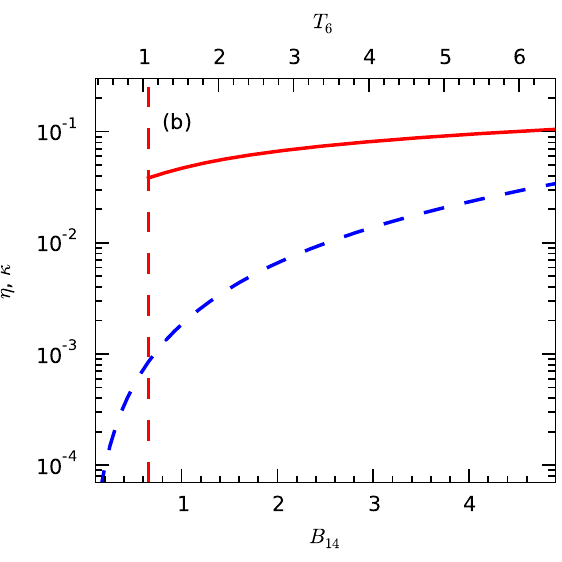}
\includegraphics{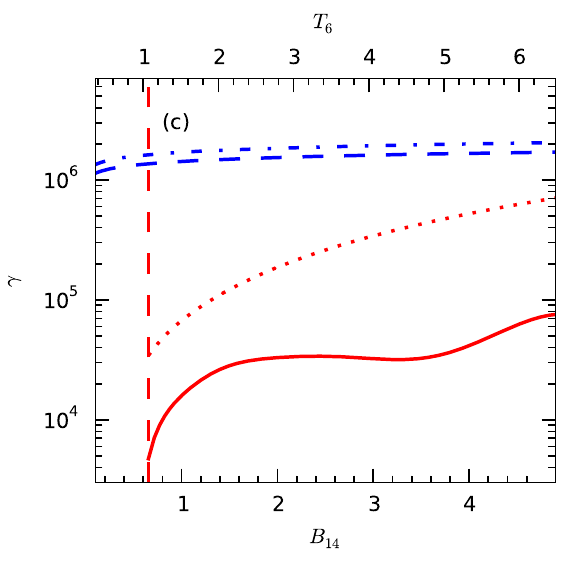} 
			\caption{Dependence of the gap height (panel a), the screening factor or the overheating parameter (panel b), and the particle Lorentz factor (panel c) on the surface magnetic field.
            Solid red lines correspond to the PSG-on mode (ICS-dominated gaps) while dashed blue lines correspond to the PSG-off mode (CR-dominated gaps).
            Calculations were performed using the following parameters: $P=0.58$, $\Re_{6}=0.6$, $B_{{\rm d}}=2.6\times10^{12}\,{\rm G}$, and $\alpha=36^{\circ}$. 
            The actual polar cap radius was calculated separately for a given surface magnetic field as $R_{{\rm pc}}=R_{{\rm dp}}\sqrt{B_{{\rm d}}/B_{{\rm s}}}$.
            In panel (c) the red solid and dotted lines correspond to characteristic and maximum Lorentz factors ($\gamma_{{\rm c}}$, $\gamma_{{\rm max}}$) in the PSG-on mode while blue dashed and dashed-dotted lines correspond to $\gamma_{{\rm c}}$ and $\gamma_{{\rm max}}$ in the PSG-off mode.
            Corresponding critical temperature is shown on the top axis of the figures.
            \label{fig:h_b14} }
		\end{figure*}

\begin{figure*}
			\centering{} \includegraphics{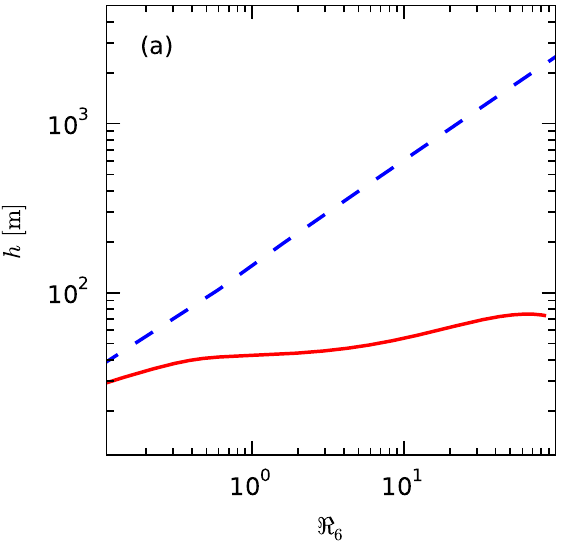} \includegraphics[width=5.7cm, height=5.5cm]{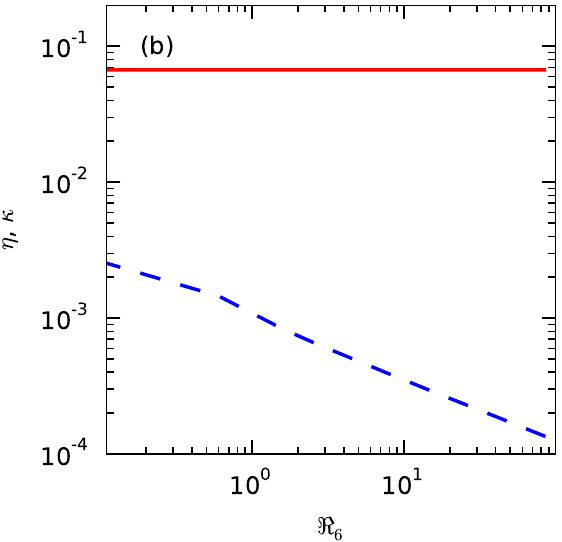} \includegraphics[width=5.7cm, height=5.5cm]{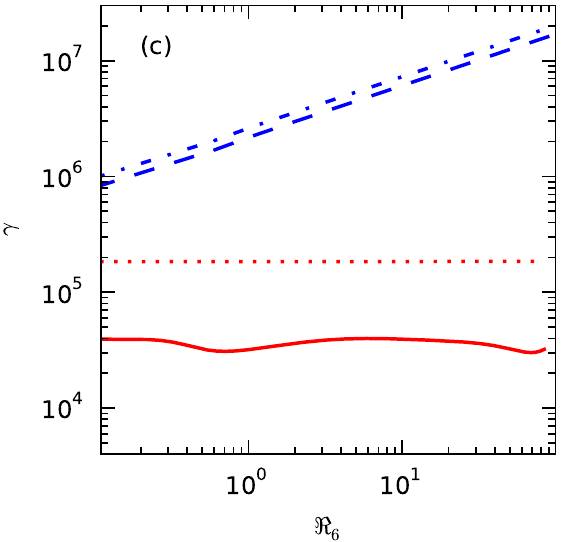} 
            \caption{Dependence of the gap height (panel a), the screening factor or the overheating parameter (panel b), and the particle Lorentz factor (panel c) on the curvature radius of magnetic field lines. 
            Calculations were performed using the following parameters: $P=0.58$, $B_{{\rm d}}=2.6\times10^{12}\,{\rm G}$, $B_{{\rm s}}=2.0\times10^{14}\,{\rm G}$, $T_{\rm s} = 3.3 \, {\rm MK}$, $\alpha=36^{\circ}$. For a more detailed description see Figure \ref{fig:h_b14}. 
            \label{fig:h_re6} }
		\end{figure*}

\begin{figure*}
			\centering{}\includegraphics{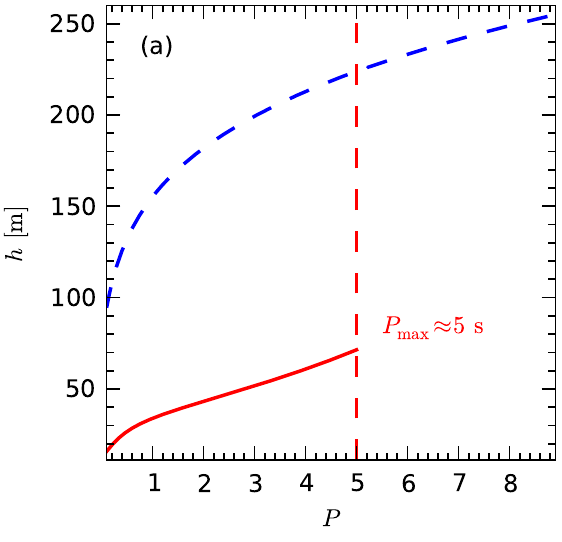} \includegraphics{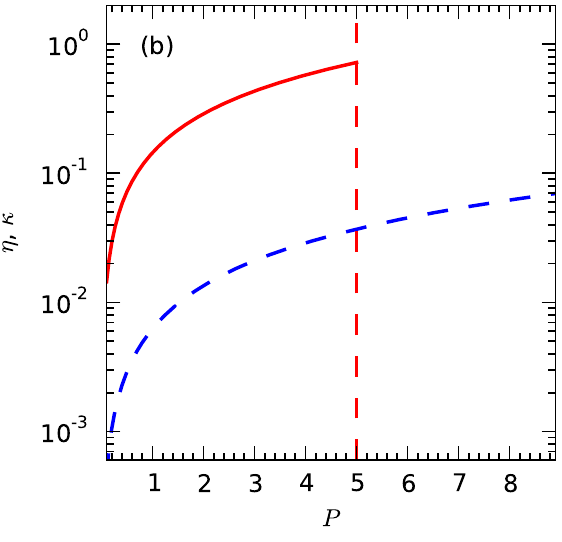} \includegraphics{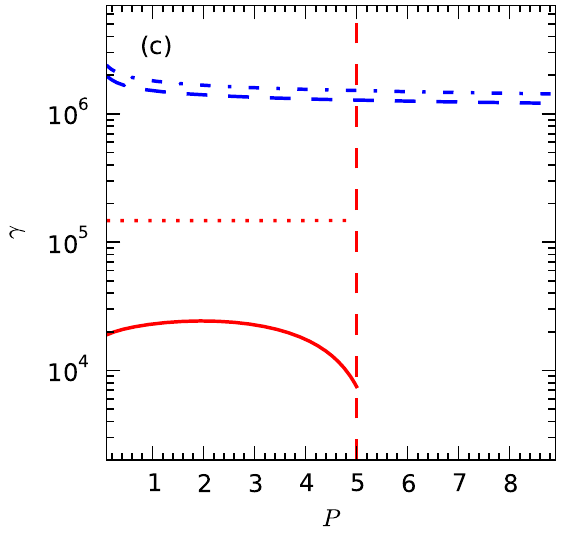} 
            \caption{Dependence of the gap height (panel a), the screening factor or a overheating parameter (panel b), and the particle Lorentz factor (panel c) on the pulsar period.
			Calculations were performed using the following parameters: $B_{{\rm d}}=2.6\times10^{12}\,{\rm G}$, $B_{{\rm s}}=2.0\times10^{14}\,{\rm G}$, $T_{\rm s} = 3.3 \, {\rm MK}$, $\alpha=36^{\circ}$, $\Re_{6}=0.3$. 
            For a given pulsar period, the period derivative was set to satisfy $B_{{\rm d}}=2.6\times10^{12}\,{\rm G}$. 
            For a more detailed description see Figure \ref{fig:h_b14}.
            \label{fig:h_p} }
		\end{figure*}

    In Section \ref{sec:results} we show possible solutions of the gap breakdown for a given pulsar parameters ($P$, $\dot{P}$, $\alpha$, $B_{\rm s}, T_{\rm s}$, $\Re$).
    We have found that for some ranges of these parameters (see Figures \ref{fig:h_b14} and \ref{fig:h_p}),
    %, for instance $B_{\rm s}<0.7\times 10^{14}\,{\rm G}$, $P> 5\, {\rm s}$ 
    the gap can operate in one mode only.
    However, there is a wide range of parameters for which the gap can operate in both the PSG-on (ICS-dominated) and PSG-off (CR-dominated) modes.
    Which of these two processes dominates the gap breakdown in this case depends on the mean free path of the primary particles $l_{\rm p}$ and the acceleration path $l_{\rm acc}$ (see Figure \ref{fig:le_gamma}).
    If $l_{\rm ICS}<l_{\rm acc}$, i.e. the particle emits ICS-photon before its energy becomes too high for efficient scattering ($\gamma \gtrsim 10^5$, see panel b in Figure \ref{fig:le_gamma}) the gap works in the PSG-on mode.
    In most cases the dominating process is the one that results in the smallest gap height, thus the ICS process is the dominating one.
    %However, it is worth noting that $l_{\rm acc}$ may be also influenced by the dynamics of spark evolution.
    %For some pulsars the ICS mode cannot be possible due to, for instance, too low screening by iron ions and, thus, too high electric field (small $l_{\rm acc}$).
    However, it is worth noting that for some pulsars the dynamics of spark evolution can influence $l_{\rm acc}$, e.g. the screening by iron ions is low and the electric field is too high to allow ICS.
    Therefore, for some pulsar parameters it could be possible that depending on details of the spark evolution the pulsar switches between the two modes.
    %Therefore, for some pulsar parameters it is difficult to ultimately answer the question in which mode the gap operates without studying the dynamics of the spark evolution.
    Finding the mechanism responsible for mode changing could explain important observations of the Chameleon pulsar \citep[PSR B0943+10,][]{2013_Hermsen, 2013_Mereghetti}, however, it is beyond the scope of this paper.
   
 	% calculated using ICSCascade (particles tracking based on magnetic field simulations,  sets:500-526), see radiation/gap.py
    
 	% PSG-on mode calculated using ICSCascade (particles tracking based on magnetic field simulations,  sets:402, 401, 400, 403, 404, 405, 406, 407), see radiation/gap.py (image size 5.7x5.5cm)
    
	% PSG-on mode calculated using ICSCascade (particles tracking based on magnetic field simulations,  sets:402, 401, 400, 403, 404, 405, 406, 407), see radiation/gap.py (image size 5.7x5.5cm)
    
	\section{Summary and discussion} \label{sec:discussion}
	
In present paper we have used a simplified approach to describe passage of the primary particles through the inner acceleration region within the framework of the PSG model.
We have identified two modes, in which the gap can operate, namely the PSG-off and PSG-on modes. 
In the PSG-off mode the acceleration potential in the gap region is high enough to enable efficient $\gamma$-photon emission by CR.
Primary particles are accelerated up to $\gamma \gtrsim 10^6$ and the gap breaks due to heating of the polar cap surface up to the critical temperature. 
Thus, the acceleration potential at the instant of gap breakdown is screened by a mixture of newly created particles and ions extracted from the stellar surface. 
The density of plasma required to overheat the polar cap is considerably smaller than the co-rotational charge density $\rho_{\rm c} \ll \rho_{\rm GJ}$ ($\kappa=10^{-3}-10^{-1}$). 
Since the magnetic field just above the stellar surface is highly non-dipolar and Lorentz factor of primary particles is  $\gamma \gtrsim 10^6$, the secondary plasma is produced mainly above the acceleration region. 
Such a scenario results in enhanced synchrotron radiation in the inner magnetosphere of a pulsar and may be associated with the increase of the non-thermal X-ray emission.
Note that despite the fact that  $\rho_{\rm c} \ll \rho_{\rm GJ}$ at the moment of gap breakdown, pair creation above the acceleration region may lead to production of plasma with density significantly exceeding the co-rotational charge density.
	
The PSG-on mode corresponds to solutions in which ICS is responsible for $\gamma$-photon emission.
Since ICS is relevant for particles with $\gamma \approx 10^3-10^4$ (see Figure \ref{fig:le_gamma}), the acceleration potential has to be reduced in order to operate in the PSG-on mode.
One of the factors which affects ICS is the density of X-ray background photons, which is the highest just above the polar cap ($z<10\,{\rm m}$).
Thus, ICS photons, and therefore pairs, are created mainly in the lower part of the gap leading to continuous heating of the stellar surface (see Figures \ref{fig:ics_sol} and \ref{fig:schematic_ics_cr}).           
The temperature of the polar cap should be maintained close to the critical temperature  leading to the thermionic emission of iron ions.
The outflow of iron ions is responsible for partial screening of the acceleration potential making the ICS process possible.
Due to thermostatic regulation (see Figure \ref{fig:thermostat}) the temperature of the polar cap is in quasi equilibrium with the backstreaming particles and the gap can break only due to production of dense enough plasma $\rho_+ + \rho_- \gg \rho_{\rm GJ}$.
In order to get the final pair multiplicity in either mode detailed full cascade simulations with the inclusion of ion extraction are necessary.
    
It was believed that formation of the inner acceleration region is strictly dependent on the high curvature of the magnetic field lines ($\Re \sim 10^6$).
However, as we have shown in this paper (see Figure \ref{fig:h_re6}) it is not the case for gaps working in the PSG-on mode. 
Even for a relatively high radius of curvature, $\Re \sim 10^8$, the height of the gap dominated by ICS is still suitable to produce radio emission ($h\lesssim 100 \, {\rm m}$).
Lower heights of ICS dominated gaps comparing with heights of CR dominated gaps is related to much higher energies of $\gamma$-photons produced by the former ($\gamma_{\rm sec}^{\rm ICS}\approx 10^3-10^4$ vs.  $\gamma_{\rm sec}^{\rm CR} \approx 10^2$).
In the accompanying paper \citep{2014_Szary_magnetars} we apply the PSG model to explain the radio emission of magnetars.

\cite{2010_Timokhin} applied the model of \cite{1975_Ruderman} to one dimensional self-consistent kinetic simulations of electron-positron cascades. 
The calculations took into account particle acceleration, pair creation and screening of the electric field by newly created pairs.
 \cite{2010_Timokhin} found that the backstreaming particles would be responsible for heating of the polar cap to temperature $T_{\rm s} \sim 4 \times 10^6 f^{-0.25}\,{\rm K}$, where $(R_{\rm dp}/c)f$ is the time between two successive discharges.
According to the calculations of the cohesive energy \citep{2007_Medin} for such a high temperatures of the polar cap the formation of the vacuum gap is not possible due to thermal ejection of iron ions.
In this paper we present more realistic model of the inner acceleration region, namely the PSG model, which allows to use the X-ray information (e.g. the polar cap size and its temperature) to estimate parameters of the acceleration gap (e.g., the gap height, the acceleration potential drop, the spark half-width, the ions density in the gap, the characteristic Lorentz factor of primary particles, etc.).
 
\section*{Acknowledgements}

This work is supported by National Science Centre Poland under grants 2011/03/N/ST9/00669 and DEC-2012/05/B/ST9/03924. We thank the anonymous referee for very constructive comments.

\bibliographystyle{mn2e}
\bibliography{bibliography}

\begin{thebibliography}{}

\bibitem[\protect\citeauthoryear{{Arons}}{{Arons}}{1983}]{1983_Arons}
{Arons} J.,  1983, \apj, 266, 215

\bibitem[\protect\citeauthoryear{{Arons}}{{Arons}}{2009}]{2009_Arons}
{Arons} J.,  2009, in {Becker} W.,  ed., Astrophysics and Space Science Library
  Vol.~357 of Astrophysics and Space Science Library, {Pulsar Emission: Where
  to Go}.
p.~373

\bibitem[\protect\citeauthoryear{{Arons} \& {Scharlemann}}{{Arons} \&
  {Scharlemann}}{1979}]{1979_Arons}
{Arons} J.,  {Scharlemann} E.~T.,  1979, \apj, 231, 854

\bibitem[\protect\citeauthoryear{{Baring} \& {Harding}}{{Baring} \&
  {Harding}}{1997}]{1997_Baring}
{Baring} M.~G.,  {Harding} A.~K.,  1997, \apj, 482, 372

\bibitem[\protect\citeauthoryear{{Baring} \& {Harding}}{{Baring} \&
  {Harding}}{2001}]{2001_Baring}
{Baring} M.~G.,  {Harding} A.~K.,  2001, \apj, 547, 929

\bibitem[\protect\citeauthoryear{{Baring}, {Wadiasingh} \& {Gonthier}}{{Baring}
  et~al.}{2011}]{2011_Baring}
{Baring} M.~G.,  {Wadiasingh} Z.,    {Gonthier} P.~L.,  2011, \apj, 733, 61

\bibitem[\protect\citeauthoryear{{Becker}}{{Becker}}{2009}]{2009_Becker}
{Becker} W.,  2009, in {W.~Becker} ed., Astrophysics and Space Science Library
  Vol.~357 of Astrophysics and Space Science Library, {X-Ray Emission from
  Pulsars and Neutron Stars}.
p.~91

\bibitem[\protect\citeauthoryear{{Bhattacharya}, {Wijers}, {Hartman} \&
  {Verbunt}}{{Bhattacharya} et~al.}{1992}]{1992_Bhattacharya}
{Bhattacharya} D.,  {Wijers} R.~A.~M.~J.,  {Hartman} J.~W.,    {Verbunt} F.,
  1992, \aap, 254, 198

\bibitem[\protect\citeauthoryear{{Chen} \& {Ruderman}}{{Chen} \&
  {Ruderman}}{1993}]{1993_Chen}
{Chen} K.,  {Ruderman} M.,  1993, \apj, 402, 264

\bibitem[\protect\citeauthoryear{{Cheng} \& {Ruderman}}{{Cheng} \&
  {Ruderman}}{1980}]{1980_Cheng}
{Cheng} A.~F.,  {Ruderman} M.~A.,  1980, \apj, 235, 576

\bibitem[\protect\citeauthoryear{{Erber}}{{Erber}}{1966}]{1966_Erber}
{Erber} T.,  1966, Reviews of Modern Physics, 38, 626

\bibitem[\protect\citeauthoryear{{Faucher-Gigu{\`e}re} \&
  {Kaspi}}{{Faucher-Gigu{\`e}re} \& {Kaspi}}{2006}]{2006_Faucher}
{Faucher-Gigu{\`e}re} C.-A.,  {Kaspi} V.~M.,  2006, \apj, 643, 332

\bibitem[\protect\citeauthoryear{{Geppert}, {Gil} \& {Melikidze}}{{Geppert}
  et~al.}{2013}]{2013_Geppert}
{Geppert} U.,  {Gil} J.,    {Melikidze} G.,  2013, \mnras, 435, 3262

\bibitem[\protect\citeauthoryear{{Gil}, {Melikidze} \& {Zhang}}{{Gil}
  et~al.}{2007}]{2007_Gil_b}
{Gil} J.,  {Melikidze} G.,    {Zhang} B.,  2007, \mnras, 376, L67

\bibitem[\protect\citeauthoryear{{Gil} \& {Melikidze}}{{Gil} \&
  {Melikidze}}{2002}]{2002_Gil_b}
{Gil} J.,  {Melikidze} G.~I.,  2002, \apj, 577, 909

\bibitem[\protect\citeauthoryear{{Gil}, {Melikidze} \& {Geppert}}{{Gil}
  et~al.}{2003}]{2003_Gil}
{Gil} J.,  {Melikidze} G.~I.,    {Geppert} U.,  2003, \aap, 407, 315

\bibitem[\protect\citeauthoryear{{Gil}, {Melikidze} \& {Mitra}}{{Gil}
  et~al.}{2002}]{2002_Gil}
{Gil} J.~A.,  {Melikidze} G.~I.,    {Mitra} D.,  2002, \aap, 388, 235

\bibitem[\protect\citeauthoryear{{Goldreich} \& {Julian}}{{Goldreich} \&
  {Julian}}{1969}]{1969_Goldreich}
{Goldreich} P.,  {Julian} W.~H.,  1969, \apj, 157, 869

\bibitem[\protect\citeauthoryear{{Gonthier}, {Harding}, {Baring}, {Costello} \&
  {Mercer}}{{Gonthier} et~al.}{2000}]{2000_Gonthier}
{Gonthier} P.~L.,  {Harding} A.~K.,  {Baring} M.~G.,  {Costello} R.~M.,
  {Mercer} C.~L.,  2000, \apj, 540, 907

\bibitem[\protect\citeauthoryear{{Hermsen}, {Hessels}, {Kuiper}, {van Leeuwen}
  \& {Mitra} D.}{{Hermsen} et~al.}{2013}]{2013_Hermsen}
{Hermsen} W.,  {Hessels} J.~W.~T.,  {Kuiper} L.,  {van Leeuwen} J.,    {Mitra}
  D. e.~a.,  2013, Science, 339, 436

\bibitem[\protect\citeauthoryear{{Jones}}{{Jones}}{1980}]{1980_Jones}
{Jones} P.~B.,  1980, \apj, 236, 661

\bibitem[\protect\citeauthoryear{{Lai}}{{Lai}}{2001}]{2001_Lai}
{Lai} D.,  2001, Reviews of Modern Physics, 73, 629

\bibitem[\protect\citeauthoryear{{Medin} \& {Lai}}{{Medin} \&
  {Lai}}{2006a}]{2006_Medin_a}
{Medin} Z.,  {Lai} D.,  2006a, \pra, 74, 062507

\bibitem[\protect\citeauthoryear{{Medin} \& {Lai}}{{Medin} \&
  {Lai}}{2006b}]{2006_Medin_b}
{Medin} Z.,  {Lai} D.,  2006b, \pra, 74, 062508

\bibitem[\protect\citeauthoryear{{Medin} \& {Lai}}{{Medin} \&
  {Lai}}{2007}]{2007_Medin}
{Medin} Z.,  {Lai} D.,  2007, \mnras, 382, 1833

\bibitem[\protect\citeauthoryear{{Medin} \& {Lai}}{{Medin} \&
  {Lai}}{2010}]{2010_Medin}
{Medin} Z.,  {Lai} D.,  2010, \mnras, 406, 1379

\bibitem[\protect\citeauthoryear{{Mereghetti}, {Tiengo}, {Esposito} \&
  {Turolla}}{{Mereghetti} et~al.}{2013}]{2013_Mereghetti}
{Mereghetti} S.,  {Tiengo} A.,  {Esposito} P.,    {Turolla} R.,  2013, \mnras,
  435, 2568

\bibitem[\protect\citeauthoryear{{Ruderman} \& {Sutherland}}{{Ruderman} \&
  {Sutherland}}{1975}]{1975_Ruderman}
{Ruderman} M.~A.,  {Sutherland} P.~G.,  1975, \apj, 196, 51

\bibitem[\protect\citeauthoryear{{Sturrock}}{{Sturrock}}{1971}]{1971_Sturrock}
{Sturrock} P.~A.,  1971, \apj, 164, 529

\bibitem[\protect\citeauthoryear{{Szary}}{{Szary}}{2013}]{2013_Szary}
{Szary} A.,  2013, PhD thesis

\bibitem[\protect\citeauthoryear{{Szary}, {Melikidze} \& {Gil}}{{Szary}
  et~al.}{2014a}]{2014_Szary_magnetars}
{Szary} A.,  {Melikidze} G.~I.,    {Gil} J.,  2014a, {On the origin of radio
  emission from magnetars}, submitted to ApJ

\bibitem[\protect\citeauthoryear{{Szary}, {Zhang}, {Melikidze}, {Gil} \&
  {Xu}}{{Szary} et~al.}{2014b}]{2014_Szary}
{Szary} A.,  {Zhang} B.,  {Melikidze} G.~I.,  {Gil} J.,    {Xu} R.-X.,  2014b,
  \apj, 784, 59

\bibitem[\protect\citeauthoryear{{Timokhin}}{{Timokhin}}{2006}]{2006_Timokhin}
{Timokhin} A.~N.,  2006, \mnras, 368, 1055

\bibitem[\protect\citeauthoryear{{Timokhin}}{{Timokhin}}{2010}]{2010_Timokhin}
{Timokhin} A.~N.,  2010, \mnras, 408, 2092

\bibitem[\protect\citeauthoryear{{Xia}, {Qiao}, {Wu} \& {Hou}}{{Xia}
  et~al.}{1985}]{1985_Xia}
{Xia} X.~Y.,  {Qiao} G.~J.,  {Wu} X.~J.,    {Hou} Y.~Q.,  1985, \aap, 152, 93

\bibitem[\protect\citeauthoryear{{Zhang}, {Qiao} \& {Han}}{{Zhang}
  et~al.}{1997a}]{1997_Zhang_b}
{Zhang} B.,  {Qiao} G.~J.,    {Han} J.~L.,  1997a, \apj, 491, 891

\bibitem[\protect\citeauthoryear{{Zhang}, {Qiao}, {Lin} \& {Han}}{{Zhang}
  et~al.}{1997b}]{1997_Zhang_a}
{Zhang} B.,  {Qiao} G.~J.,  {Lin} W.~P.,    {Han} J.~L.,  1997b, \apj, 478, 313

\end{thebibliography}
  
\label{lastpage}

\end{document}